%% file: main.tex
\documentclass[twocolumn]{aastex7}

\usepackage{graphicx}
\usepackage{amsmath}
\usepackage[dvipsnames]{xcolor}

\renewcommand{\arraystretch}{1.2}  
\usepackage{listings}
\usepackage{color}
\definecolor{dkgreen}{rgb}{0,0.6,0}
\definecolor{gray}{rgb}{0.5,0.5,0.5}
\definecolor{mauve}{rgb}{0.58,0,0.82}
\definecolor{golden}{rgb}{0.86,0.65,0.01}
\usepackage{color}
\usepackage{soul}
\usepackage{amsmath}
\usepackage{xspace}
\usepackage{graphicx}
\usepackage{enumitem}
\usepackage{amssymb}
\usepackage{xifthen}
\usepackage{hyperref}
\usepackage[normalem]{ulem}
\usepackage{multirow}
\usepackage{threeparttable}
\usepackage{placeins}
\usepackage{tabularx}
\usepackage{makecell}

\newcommand{\gaia}{\textit{Gaia}\xspace}

\begin{document}

\title{The DECam MAGIC Survey: Uncovering the Tidal Tails of the Crater II Dwarf Galaxy}

\input{authors.tex}

\begin{abstract}
  
Crater II (CraII), a large and low-density dwarf spheroidal galaxy, has unusual observed properties that are difficult to reproduce in cold dark matter simulations.  
Ongoing tidal disruption may help explain the discrepancies, as  evidenced by the recent discovery of tidal tails.
Here we present metallicity-sensitive narrowband photometry of the Ca II H and K lines from the Dark Energy Camera, covering $128$ deg$^2$ across the center and identified tidal tails of CraII as part of the Mapping the Ancient Galaxy in CaHK (MAGIC) survey.
Our combined photometric metallicity, color-magnitude, proper motion, and parallax selections identify 162 CraII candidates. 
Of these, 37 candidates are located in the tidal tails which extend at least $7^\circ$ ($\sim 95$ kpc) from the center of CraII, suggesting it has lost $\gtrsim 25\%$ of its initial stellar mass. 
We confirm low contamination rates with dedicated control fields and highlight the extremely low surface brightness stellar features that can be uncovered with CaHK data, as faint as $\sim 36$ mag arcsec$^{-2}$.
We also make the first detection of a metallicity gradient ($-0.34\pm0.17~{\rm dex}~{\rm deg}^{-1}$) in the center of the galaxy and infer a stream width of $w\sim 0.8^\circ$, roughly 50\% larger than the CraII half-light radius.
The detection of candidates in the most distant CraII pointings from its center implies that the tidal tails extend beyond our footprint.  
We compare the CraII stream to $N$-body models with ``cored" and ``cuspy" dark matter halo progenitors, determining that CraII's density profile is still ambiguous and warrants further modeling.

\end{abstract}


\section{Introduction}

Dwarf galaxies \citep[$M_V \gtrsim -18,$][]{Grebel03,McConnachie12} are the most common type of galaxy in the Universe \citep{Marzke97} and the Milky Way (MW) provides a nearby useful sample of dwarf spheroidal galaxies (dSphs).
MW dSphs are low-mass, dark matter-dominated systems that probe Lambda cold dark matter ($\Lambda$CDM) and preserve some of the oldest known stellar populations with the lowest measured metallicities \citep{Mateo98,Tolstoy2009ARA&A..47..371T, Brown12,Frebel12,Frebel15,Ji15,Chiti25}.
Their faint and extended nature makes them difficult to discover without reliable, high fidelity photometry.

Modern wide-area digital sky surveys have detected numerous dwarf galaxies \citep[e.g.,][]{Koposov08,Richardson11} and continue to detect numerous new faint MW systems \citep[e.g.,][]{Willman05,Belokurov07a,Bechtol15,Cerny23,Smith24,Tan25}.
High-precision spectroscopic radial velocities of individual stars in these systems offer us insight into the internal kinematics of faint dwarfs \citep{1983ApJ...266L..11A,Mateo94,Olszewski98,Simon07,Torrealba19}.
Due to their shallower dark matter potentials, dSphs are also highly sensitive to mechanisms such as feedback through star formation, reionization, ram pressure stripping, and tidal stripping \citep{McConnachie12,2021ApJ...913...53P,Sales22, 2022NatAs...6..647C}.

Tidal interactions between dwarf galaxies and the MW can cause them to grow in radius and reduce their velocity dispersion \citep[e.g.,][]{2008ApJ...673..226P, Errani15}.
Stars are lost from the dwarf galaxy to leading and trailing orbits, filling a larger volume than the original system.
Over multiple orbits, these tidal interactions can become strong enough to produce coherent stellar streams before phase mixing into the stellar halo \citep{Helmi99}.
These streams can extend over a long portion of an orbit, with only a small part of the progenitor's original stellar mass remaining when tidal disruption is nearly complete \citep{2008ApJ...673..226P}.
Although galaxy formation models predict large numbers of globular cluster and dwarf galaxy stellar streams in the MW’s halo from past Galactic accretion events \citep{2005ApJ...635..931B, Helmi20}, and there are over 100 known MW stellar streams \citep[e.g.,][]{Mateu2023MNRAS.520.5225M, Bonaca25}, only a handful of dwarf galaxies themselves exhibit clear tidal stripping signatures \citep[e.g. Sagittarius, Tucana III, Antlia II, Crater II, and Bo{\"o}tes III;][]{Ibata94,Majewski03,Belokurov06,Li18,2018ApJ...865....7C,Ji2021,Pace22, Coppi24}.
However, some hydrodynamical simulations predict that most satellite galaxies with orbital pericenters $\lesssim 100$ kpc contain faint stellar streams undetectable by current observational detection limits \citep{Shipp23,Riley25,Shipp25}.

The CraII dSph was discovered using ATLAS photometry from the VLT Survey Telescope \citep{Torrealba16}, and we summarize several of its observational properties in Table \ref{tab:props}.
Notably, it is similar in physical size (half-light radius $r_h \sim 1~{\rm kpc}$) to the SMC, but three orders of magnitude less luminous.
It is one of the lowest surface brightness systems in the Local Group along with Antlia II \citep{Torrealba19} and Andromeda XIX \citep{Martin16,Collins20} and may represent a local, lower luminosity counterpart of ultra-diffuse galaxies \citep[e.g.,][]{Koda15,vanDokkum15,Lim20,Gannon24}.
The velocity dispersion of CraII is among the lowest of the dSphs, but  still implies a large dynamical mass-to-light ratio ($\sim 30$) and significant dark matter halo mass.
These unusual properties suggest that the system likely formed in an irregular way; possibly in an uncommonly low density dark matter halo \citep{Amorisco19} or from severe Galactic tides \citep[e.g.,][]{Frings17,Applebaum21}.

Previous studies using \gaia astrometry and AAT spectroscopy \citep{Ji2021,Limberg25} suggest a tidal origin given CraII’s large size and low density. 
Its position on the luminosity-metallicity relation suggests a maximum 42\% mass loss (see Figure 8 of \citealt{Ji2021}). 
However, this is difficult to reconcile with the \textgreater~90\% predicted mass loss within $\Lambda$CDM cosmology from $N$-body models \citep{Sanders18,Borukhovetskaya22}.
Intrinsic scatter in the mass-metallicity relation may address this mismatch \citep[e.g.,][]{Williamson16,Riley25b}, but this remains an open question.
\citet{Sanders18} find that it is difficult to simultaneously simulate CraII's large size and low velocity dispersion in ``cuspy" (density $\rho(r) \propto r^{-1}$) dark matter halos.
It is also unlikely that CraII is far from dynamical equilibrium because it is far from its pericenter and the core should have returned to equilibrium \citep{Fu19,Borukhovetskaya22}.

Therefore, CraII's properties can be more easily explained if its dark matter halo has been ``cored" ($\rho(r < r_{\rm core}) \approx \rho_0$) through a mechanism such as stellar feedback or alternative dark matter physics, making them less dense at the center and more easily stripped by tides.
This is a phenomenon recognized but poorly understood in dwarf galaxies \citep{Strigari17,Genina18,Harvey18,Sales22}.
Based on the radial velocity and velocity dispersion profiles of their member stars, many studies have argued that dSphs may have cored rather than the cuspy density profiles predicted in $\Lambda$CDM \citep{Battaglia08,Walker11,Amorisco12}.
The degree of core formation is known to increase with number of starburst cycles on long cosmological timescales, suggesting dSphs with more extended star-formation histories should have more prominent cores \citep{Onorbe15,Chan15,ElBadry16,Read17}.
\citet{Walker19} discovered that CraII experienced multiple early star formation events from a color-magnitude analysis, but the duration of its star formation history ($\sim 2$ Gyr) is likely too short to produce a core.

Recent evidence of a stellar stream attached to the central remnant of CraII strongly argues that the system is experiencing tidal stripping \citep{Limberg25}.
This has also been previously inferred from RR Lyrae (RRL) measurements that trace a linear distribution on the sky across the expected tidal tails \citep{Coppi24,Vivas25}.
Moreover, tidal tails of a cored halo should be wider than those of a cuspy halo \citep{Errani15}, so measuring CraII’s stream width may allow us to constrain this aspect of its dark matter halo mass profile.

Here we present analysis of metallicity-sensitive narrowband photometry to identify the tidal tails of CraII.
This imaging helps to differentiate low-metallicity stars associated with CraII from more metal-rich MW contaminant stars.
Such imaging has been a powerful tool in identifying the faint low-metallicity outskirts of dwarf galaxies \citep[e.g.,][]{Chiti21,Longeard22,Longeard23,Ou24,Chiti25}.
Section \ref{sec:data} describes our new observations and supplementary data from survey programs.
This data is compared to theoretical modeling and transformed to stream coordinates, which we introduce in Section \ref{sec:modeling}.
Sections \ref{sec:methods} and \ref{sec:membership} explain our analysis for selecting CraII stars from the candidates and then present the resulting candidates across the footprint, respectively.
We discuss implications for tidal disruption and dark matter modeling of CraII in Section \ref{sec:discussion}.
Finally, we summarize our findings and suggest follow-up observations in Section \ref{sec:conclusion}.

\begin{deluxetable}{ccc}
\tablecolumns{3}
\tablecaption{\label{tab:props}Crater II Properties}
\tablehead{Parameter & Value & Reference}
\startdata
$\alpha$ (deg; J2000) & 177.310 & T16 \\
$\delta$ (deg; J2000) & $-18.413$ & T16 \\
$\text{M}_\star$ ($\text{M}_\odot$) & $10^{5.55}$ & J21 \\
$M_V$ & $-8.18\pm0.10$ & T16 \\ 
$r_h$ (arcmin) & $31.2~\pm~2.5$ & T16 \\
$r_h$ (pc) & $1050~\pm~90$ & T16 \\
$(m-M)_0$ & $20.333\pm0.068$ & V20 \\
$d~{\rm (kpc)}$ & $116.6_{-3.6}^{+3.7}$ & V20 \\ 
$\mu_{r_h}$ (mag arcsec$^{-2}$) & $30.6~\pm~0.2$ & T16 \\
$\sigma_v$ (km s$^{-1}$) & $2.34^{+0.42}_{-0.30}$ & J21 \\
$r_{\rm peri}$ (kpc) & $24.0^{+5.6}_{-5.2}$ & P22 \\
$r_{\rm apo}$ (kpc) & $138.1^{+7.9}_{-4.9}$ & P22 \\
$\overline{\mu}_{\alpha\star}$ (mas yr$^{-1}$) & $-0.072^{+0.020}_{-0.020}$ & P22 \\
$\overline{\mu}_{\delta}$ (mas yr$^{-1}$) & $-0.112^{+0.013}_{-0.013}$ & P22 \\
${\rm [Fe/H]}$ (dex) & $-2.16\pm0.04$ & J21 \\
$\sigma_{\rm [Fe/H]}$ (dex) & $0.24\pm0.05$ & J21 \\
\enddata
\tablecomments{References: (T16) \citet{Torrealba16}; (J21) \citet{Ji2021}; (V20) \citet{Vivas20}; (P22) \citet{Pace22}}
\end{deluxetable}

\section{Data}\label{sec:data}

Our data combines new metallicity-sensitive narrowband imaging with existing broadband coverage over $\sim 100 {\rm ~deg}^2$.
We first introduce the design of our observational CraII program.
Then, we review our sources of additional imaging and analyze their application to this study.

\begin{figure*}
    \centering
    \includegraphics[width=1\linewidth]{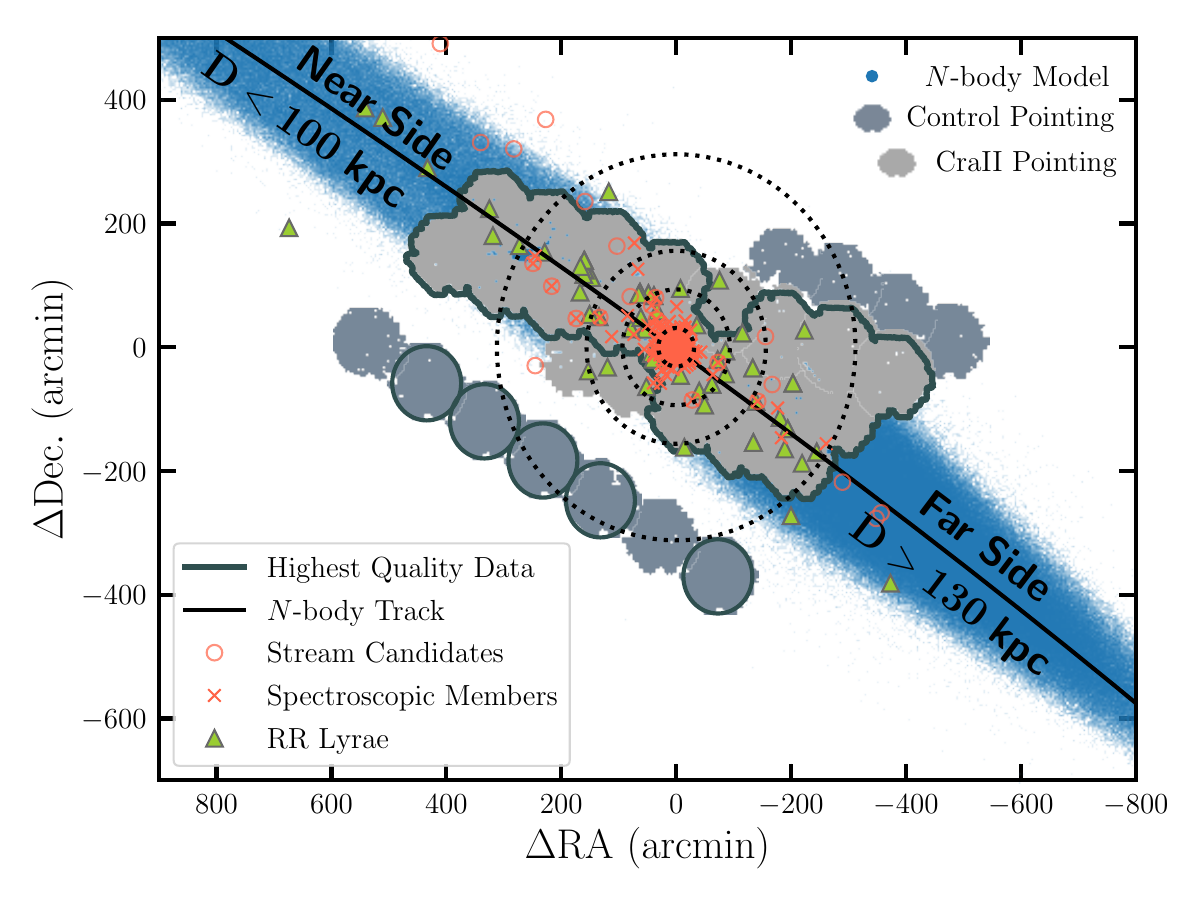}
    \caption{Twenty-one observed DECam pointings (gray) and eleven candidate control fields (blue-gray) relative to the center of CraII.
    The direction of the stream track as predicted by the $N$-body model in blue (see Section \ref{sec:$N$-body}) is shown as a solid black line.
    Pointings within the dark teal outlines were observed under the best weather conditions and dithered, providing the highest quality data of the sample.
    The 2 fields outside the outline but directly next to the CraII core were dithered in poor conditions and the remaining fields were not dithered.
    The 25 dithered pointings are 1 magnitude deeper than these 7. 
    Concentric 1, 3, 5, and 10 half-light radii \citep{Torrealba16} are drawn as dotted black circles.
    RGB stream candidates from \citet{Coppi24} are marked as orange circles and known spectroscopic members \citep{Ji2021,Limberg25} are marked as orange Xs.
    CraII stream RRL from \citet{Coppi24} and \citet{Vivas25} are shown as green triangles.
    }
    \label{fig:CraII_pointing}
\end{figure*}

\subsection{DECam N395 Imaging of Crater II}\label{sec:N395}

We use the Dark Energy Camera (DECam; \citealt{Flaugher15}) instrument's N395 filter on the V\'ictor M. Blanco 4-meter telescope at the NSF Cerro Tololo Inter-American Observatory to collect photometry from 21 fields surrounding CraII and 11 control fields for comparison as seen in Figure \ref{fig:CraII_pointing}.
These were chosen to follow the orientation of the stream track from an $N$-body model (see Section \ref{sec:$N$-body}) and cover its predicted width.
Pointings within the highlighted polygons were observed under the best weather conditions (PSF $\lesssim 1\farcs15$) and dithered ($3\times12$ min), providing the highest quality data of the sample.
The remaining fields were either taken during poorer weather or were not dithered.

\setcounter{footnote}{0}
The majority of our data were obtained from 4.5 nights of dedicated CraII programs (NOIRLab Prop. ID 2024A-974884; PI: A. B. Pace, and NOIRLab Prop. ID 2025A-402104; PI: A. B. Pace).
The rest of the data was obtained as part of the MAGIC survey (NOIRLab Prop. ID 2023B-646244; PI: A. Chiti) when the MAGIC footprint was not observable.
Figure \ref{fig:CraII_pointing} shows the $128~{\rm deg}^2$ footprint obtained with multiples of 12 minute exposures in gray that extends more than 15 half-light radii from the center of CraII where there were known spectroscopic members at the time of our observations \citep{Ji2021}.
Recent spectroscopic members discovered by the Southern Stellar Stream Spectroscopic Survey ($S^5$) are also included \citep{Li19,Limberg25}.
Our exposure time was set to reach S/N $\sim33$ at $g\sim 20.5$. This reaches the upper red giant branch (RGB) stars in the center of CraII in the magnitude range of $g\sim 20.5-17$. 
This magnitude limit further corresponds to where {\it Gaia} DR3 astrometry starts to lose its constraining power.
RRL stars have also been detected around CraII \citep{Vivas20,Coppi24,Vivas25}, and the RRL stars across the tidal tails in \citet{Coppi24} and \citet{Vivas25} are shown in green in Figure~\ref{fig:CraII_pointing}.

The N395 narrow-band filter has a central wavelength of 3952 \AA, a bandpass of 10 nm, and a ``top hat" transmission profile that covers the \ion{Ca}{2} H and K (CaHK) lines at 3968.5 \AA~and 3933.7 \AA, respectively. 
CaHK narrowband photometry allows reliable metallicity discrimination of faint RGB stars, enabling unprecedented sensitivity to ultra-diffuse tidal features.
Hereafter, we refer to the de-reddened photometric magnitude of the N395 filter as ${\rm CaHK}_0$.
We use the MAGIC processing pipeline to photometer, zeropoint-calibrate, and derive metallicities from this photometry (see Section~\ref{sec:magic}; described in detail in \citealt{Chiti26}).

\subsection{DELVE}

Our new observations are also combined with broadband data from the DECam Local Volume Exploration Survey (DELVE) for the metallicity derivation in Section~\ref{sec:magic}.
DELVE (NOIRLab Prop. ID 2019A-0305; PI: A. Drlica-Wagner) leverages data from existing archival DECam programs, the Dark Energy Survey \citep{DES05,DES16}, the Dark Energy Camera Legacy Survey \citep{Dey19}, the DECam eROSITA Survey \citep{Zenteno25}, and its own 150+ nights of observations to survey the Southern Hemisphere to search for new dwarf galaxies.
DELVE DR2 \citep{Drlica-Wagner22} covered \textgreater~21000 deg$^2$ at high-Galactic-latitude ($|b|$\textgreater $10^\circ$) in four optical/near-infrared filters ($g$, $r$, $i$, $z$).
Overall, the survey aims to investigate the abundance, properties, and environment of satellites to learn about the galaxy-halo connection.
The measured flux through the N395 filter combined with DELVE's broadband coverage enables MAGIC to derive photometric metallicities.

\subsection{MAGIC Photometric Metallicities and Validation}\label{sec:magic}

\begin{figure}
    \centering
    \includegraphics[width=1\linewidth]{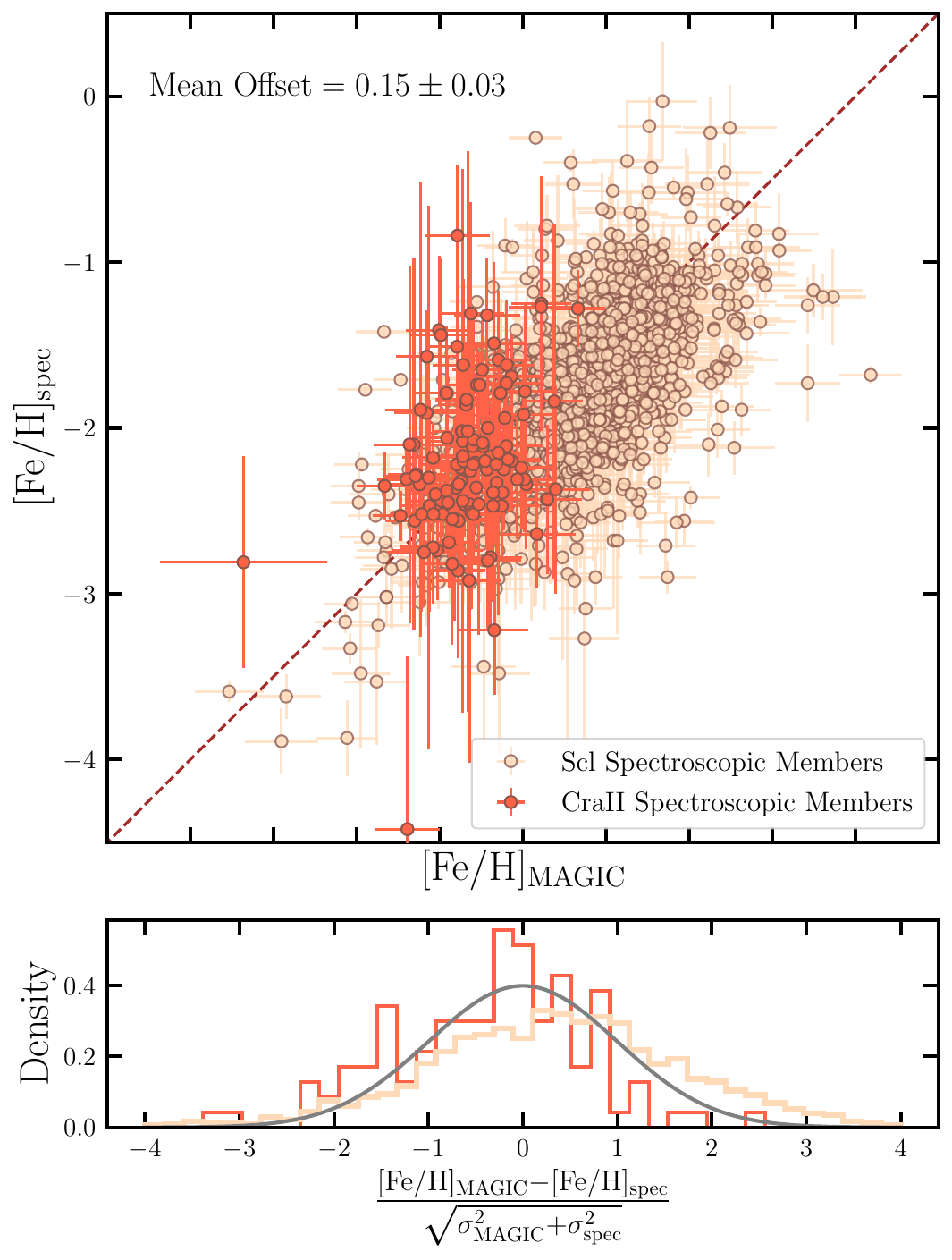}
    \caption{Comparison of 114 CraII (orange) stellar metallicities with CaHK photometry and high-resolution calcium triplet spectroscopic metallicities from the AAOmega spectrograph on the 3.9 m Anglo-Australian Telescope \citep{Ji2021}.
    We also show 1969 previously measured metallicity validation results from \citet{Barbosa25} for spectroscopic measurements \citep{Reyes22,Tolstoy23} of stars in the Sculptor (light orange) dwarf galaxy.
    The upper panel gives both MAGIC CaHK and spectroscopic metallicities and the bottom panel shows normalized difference histograms between estimates.
    The dashed brown line traces where the metallicities are equal and the gray curve illustrates a Gaussian function with mean = 0 and standard deviation = 1 for comparison.
    We note that this offset implies our CaHK metallicities should all shift by a small correction factor that we do not apply in this study.}
    \label{fig:met}
\end{figure}

The MAGIC survey \citep{Chiti26} is an observational program that studies the most metal-poor stars in the MW halo.
With 54 nights spread over six observing semesters, MAGIC aims to chemically map 5400 deg$^2$ of the MW and its structures using the DECam N395 filter.
We process our data using the MAGIC pipeline, which generates merged and reddening-corrected source catalogs, and derives photometric metallicities from DECam Community Pipeline \citep{Valdes14} processed CaHK images.
More information about the data acquisition, reduction, and photometering of reduced data is detailed in the survey overview and early science paper \citep{Chiti26}.

Prior work with MAGIC survey processing and analysis demonstrate its potential for our work, including a spatially unbiased study of the metallicity distribution function of the Sculptor dSph \citep{Barbosa25}, follow-up studies targeting the most metal-poor stars in the distant Milky Way
\citep{Placco25}, validating candidate members of the Pictor II ultra-faint dwarf galaxy (UFD) \citep{Pace25b}, and identifying the first ultra metal-poor star in a UFD \citep{Chiti25}.

To derive photometric metallicities, we forward-model fluxes through the CaHK, $g$, and $i$ filters using a grid of flux-calibrated synthetic spectra and use them to populate the color-color space, ${\rm CaHK}_0 - g_0 - 0.9\times(g_0-i_0)$ vs. $g_0-i_0$, for a range of effective temperatures (4000\,K $< T_{\text{eff}} <$ 8000\,K), surface gravities ($-0.5$\,dex $<$ log g $< 5.5$\,dex), and metallicities ($-5$\,dex $< $ [Fe/H] $< 1$\,dex). 
This color-color space allows one to map these color values to metallicities, closely following methodology in the literature \citep{Chiti20,Chiti21b,Barbosa25}.
This grid was generated using the Turbospectrum radiative transfer code \citep{Alvarez1998_turbospec, Plez2012_turbospec}, with MARCS model atmospheres \citep{Gustafsson2008_marcsmodel}, and the linelists generated from the Vienna Atomic Line Database (VALD; \citealt{Piskunov1995}, \citealt{Ryabchikova2015}). 
At this point, we note that CaHK filters have extensively been used for deriving photometric metallicities with a variety of approaches (e.g., empirical calibrations, neural-networks) in other studies conducted at different facilities that require different calibrations from DECam \citep{Starkenburg17,Whitten19,Whitten21,Galarza22,Huang22}.

We validate photometric metallicities from the pipeline specifically used in our study by comparing to Calcium Triplet spectroscopic metallicities of stars in CraII from \citet{Ji2021} and \citet{Limberg25}. 
They selected targets from a combined parallax, proper motion (mean CraII motion from \citealt{Fritz18}), and broadband photometric selection with the AAOmega spectrograph on the 3.9 m Anglo-Australian Telescope.
Of 354 targets observed by \citet{Ji2021}, 215 are marked as stars by their pipeline and have S/N \textgreater~3.
\citet{Limberg25} find another 25 stars in the center and tidal tails of CraII.
From these, they derive spectroscopic metallicities from the equivalent width of calcium triplet lines.
\citet{Ji2021} derive metallicities from both \texttt{rvspecfit} \citep{Koposov19b} and the \citet{Carrera13} CaT calibration, but we only use the CaT metallicities from \citet{Ji2021} because of a systematic offset in \texttt{rvspecfit} metallicities. The metallicities from \texttt{rvspecfit} in \citet{Limberg25} may appear offset from our CaHK metallicities as a result.
Figure \ref{fig:met} compares the spectroscopic and CaHK-derived metallicities of all CraII candidates (orange), and also overplots a metallicity comparison from a previous MAGIC study of the Sculptor dSph (\citealt{Barbosa25}; light orange).
Sculptor is approximately equal in mass to CraII (${\rm M}_*\sim10^6~{\rm M}_\odot$) but is more metal-rich and not tidally disrupted.
A weighted mean offset of $0.15\pm0.03$ and weighted scatter of 0.35 is present in CraII CaHK metallicities, with residuals showing a linear correlation with metallicity. 
However, these effects are sufficiently small overall that no correction is applied.
The majority of metallicities are nearly equal (along the brown dashed line) given their uncertainties, thus validating our use of CaHK metallicity.

\section{Modeling}\label{sec:modeling}

As discussed in Section \ref{sec:N395}, we used an $N$-body simulation of CraII to motivate our observational pointings.
Here we describe the details of the $N$-body simulation and the stream track that we derived from it.

\subsection{$N$-body Simulation}\label{sec:$N$-body}

In order to compare our observations with theoretical models, we simulate the disruption of CraII with an $N$-body simulation. 
This is the same \texttt{cusp-base} model described in \citet{Limberg25}.
We stress that these simulations are not tailored to match the tidal tails but are instead just based on the structural properties of the CraII dwarf and its present day phase-space coordinates. 
We perform these simulations with the $N$-body part of \textsc{gadget-3} which is similar to \textsc{gadget-2} \citep{Springel2005}. 
We model CraII as a Plummer sphere \citep{Plummer1911} embedded in a Navarro-Frenk-White (NFW) halo \citep{Navarro96} and we generate the self-consistent initial conditions with \textsc{agama} \citep{Vasiliev2019}. 
For the stars, we use a Plummer sphere with a mass of $3.16\times10^{5} ~{\rm M}_\odot$ and a scale radius of 1.066 kpc, matching CraII's stellar mass and its half-light radius. 
To avoid an overly extended Plummer sphere, we use an exponential truncation with a characteristic radius of 10 scale radii. For the dark matter halo, we use an NFW halo with a mass of $M_{200} = 10^7 ~{\rm M}_\odot$. 
We set the concentration using the mass-concentration relation from \cite{Dutton+2014}, assuming a Hubble constant of $67.9~{\rm km~s^{-1}~Mpc^{-1}}$ \citep{Planck2020}. This choice of the halo mass gives a velocity dispersion within the half-light radius of $3.14~{\rm km~s^{-1}}$, in rough agreement with the observed value \citep[$\sigma_v=2.55^{+0.33}_{-0.30}~{\rm km~s^{-1}}$;][]{Limberg25}. 
To eliminate distant dark matter particles in the NFW profile, we include an exponential truncation with a truncation radius of two virial radii.
We model this system with $10^6$ particles each for the stars and dark matter with a gravitational softening of 56.7 pc. 

We model the MW using the three-component model from \texttt{MWPotential2014} \citep{galpy}. 
We also include the effect of the LMC which has been shown to significantly affect the orbits of dwarfs and their stellar streams \citep[e.g.][]{Erkal19,Patel20,Garavito21,Ji2021,Shipp+2021}. 
We model the LMC as a Hernquist profile with a mass of $1.5\times10^{11} ~{\rm M}_\odot$ and a scale radius of 17.13 kpc chosen to match the enclosed mass of the LMC within 8.7 kpc \citep{vanderMarel+2014}. 
For the present-day phase space coordinates of the LMC we use its observed proper motions, distance, and radial velocity from \citet{Kallivayalil13}, \citet{Pietrzynski19}, and \citet{vanderMarel02}, respectively.
For the present-day phase space coordinates of CraII, we use the values in \cite{Pace25a}. 
We rewind the orbit of CraII in the combined potential of the MW and LMC for 7 Gyr \citep[e.g.,][]{Vasiliev23} and find, in addition to its recent apocentric passage, it has an apocenter 6.55 Gyr ago. 

\subsection{Stream Track}\label{sec:stream_track}

The stream track in Figure \ref{fig:CraII_pointing} maps the $N$-body model in a coordinate system defined along the longitude ($\phi_1$) and latitude ($\phi_2$) of the stream.
The center of CraII (RA,Dec.) = ($177.310^\circ$,$-18.413^\circ$) as found by \citet{Torrealba16} becomes ($\phi_1$,$\phi_2$) = ($0^\circ$,$0^\circ$).
We convert from celestial to stream coordinates using the $3\times3$ rotation matrix from \citet{Limberg25}:
\begin{align}
    R &= \begin{bmatrix}
        R_{0,0} & R_{0,1} & R_{0,2} \\
        R_{1,0} & R_{1,1} & R_{1,2} \\
        R_{2,0} & R_{2,1} & R_{2,2}
    \end{bmatrix} \nonumber \\
    &= \begin{bmatrix}
        -0.94775886 & 0.04452939 & -0.31586432 \\
        -0.19840253 & -0.85765902 & 0.47440218 \\
        -0.24977905 & 0.51228716 & 0.82168869
    \end{bmatrix}.
\end{align}
Then, we fit a two-variable cubic function to the $N$-body model in stream coordinates for the general area covered by our data ($|\phi_1|<15^\circ$).
These coordinates are convenient for measuring the stream width, as explained in Section \ref{sec:width}.

\section{Methodology}\label{sec:methods}

Here, we discuss the color-magnitude, proper motion, parallax, and metallicity selections used to identify candidates of CraII in our data.
In Section \ref{sec:gradient}, we describe how we account for the predicted distance gradient along CraII's tails when identifying distant candidates.
Then, in Section \ref{sec:mw_foreground}, we evaluate the robustness of our detection of distant CraII candidates along its stream relative to contamination  from MW foreground stars along the line-of-sight.

\subsection{Member Selection}\label{sec:selections}

\begin{figure*}
\centering
\includegraphics[width=\linewidth]{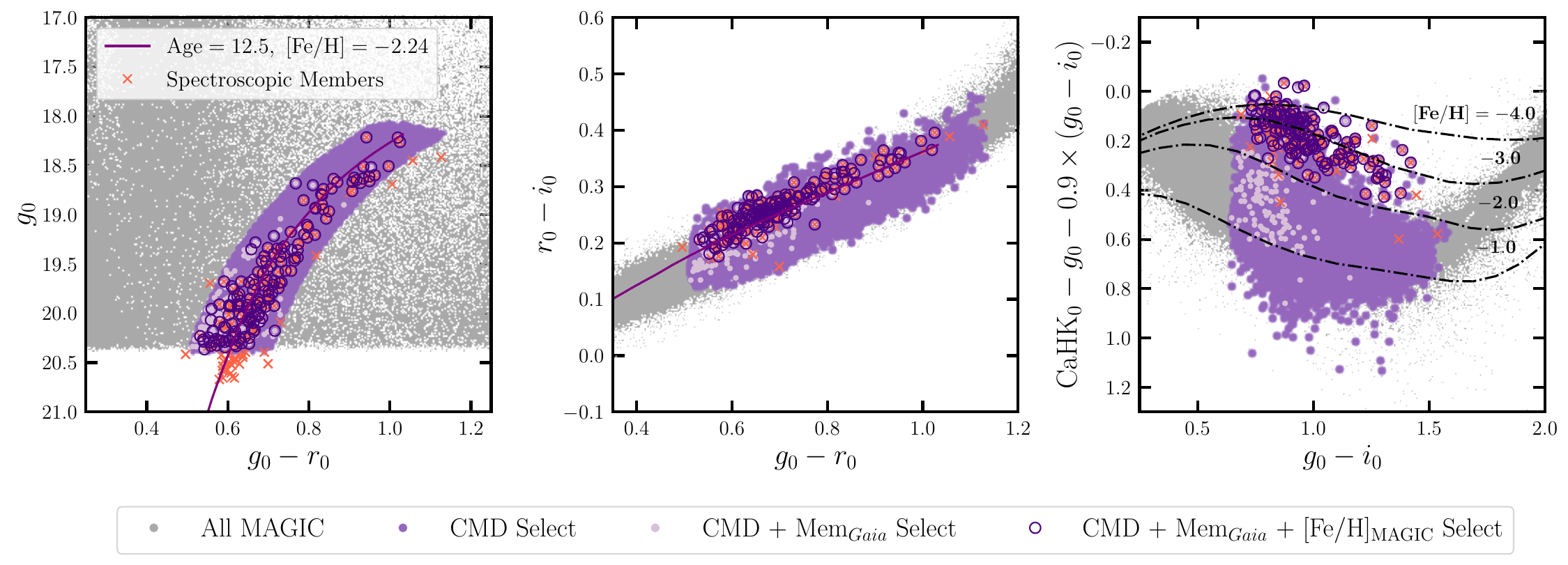}
\caption{Color-magnitude and color-color diagrams of MAGIC data after each selection.
All quality cut MAGIC data surrounding CraII (as seen in Figure \ref{fig:CraII_pointing}) is shown in gray.
The remaining data after sequential color-magnitude (purple circles), membership score (light purple circles), and metallicity (open dark purple circles) selections are further shown.
The color-magnitude cut selects stars around the isochrone in purple from the Dartmouth Stellar Evolution Database \citep{Dotter08}.
Known spectroscopic members \citep{Ji2021,Limberg25} are marked as orange Xs.
\textbf{Left:} Color-magnitude diagram with DECam $g_0$, $r_0$, $i_0$ photometry.
\textbf{Center:} Color-color diagram in $r_0-i_0$ vs $g_0-r_0$.
\textbf{Right:} Metallicity-sensitive color-color space.
The lowest metallicities are located at the top and become more metal-rich as they move down the y-axis.
The dashed-dotted curves in black are lines of fixed [Fe/H] at $\log{\rm g}=2$.
}
\label{fig:cmds}
\end{figure*}

\begin{figure*}
\centering
\includegraphics[width=\linewidth]{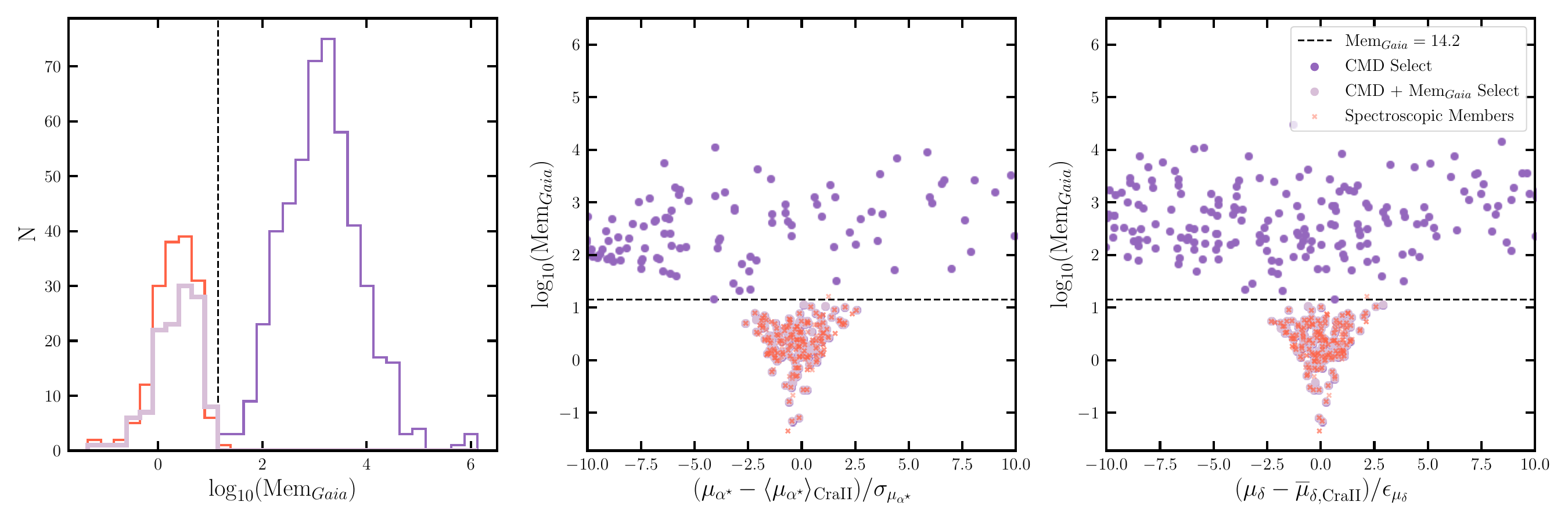}
\caption{Candidates in the central CraII pointing before and after applying proper motion and parallax criteria from the membership score \citep{Tolstoy23}.
The membership score limit, ${\rm Mem}_{Gaia} = 14.2$, is determined by $3\sigma$ for a 1D normal distribution and is drawn as a dashed black line.
The remaining data after quality cuts is sequentially selected by color-magnitude and membership score selections, shown in purple and light purple, respectively.
Known spectroscopic members \citep{Ji2021,Limberg25} are marked as orange Xs.
\textbf{Left:} Histogram of $\log_{10}({\rm Mem}_{Gaia})$ for CraII candidates and known members.
The distribution of $\log_{10}({\rm Mem}_{Gaia})$ is shown as a function of \textbf{Center:} proper motion in right ascension and \textbf{Right:} proper motion in declination.}
\label{fig:z} 
\end{figure*}

\begin{figure*}
\centering
\includegraphics[width=\linewidth]{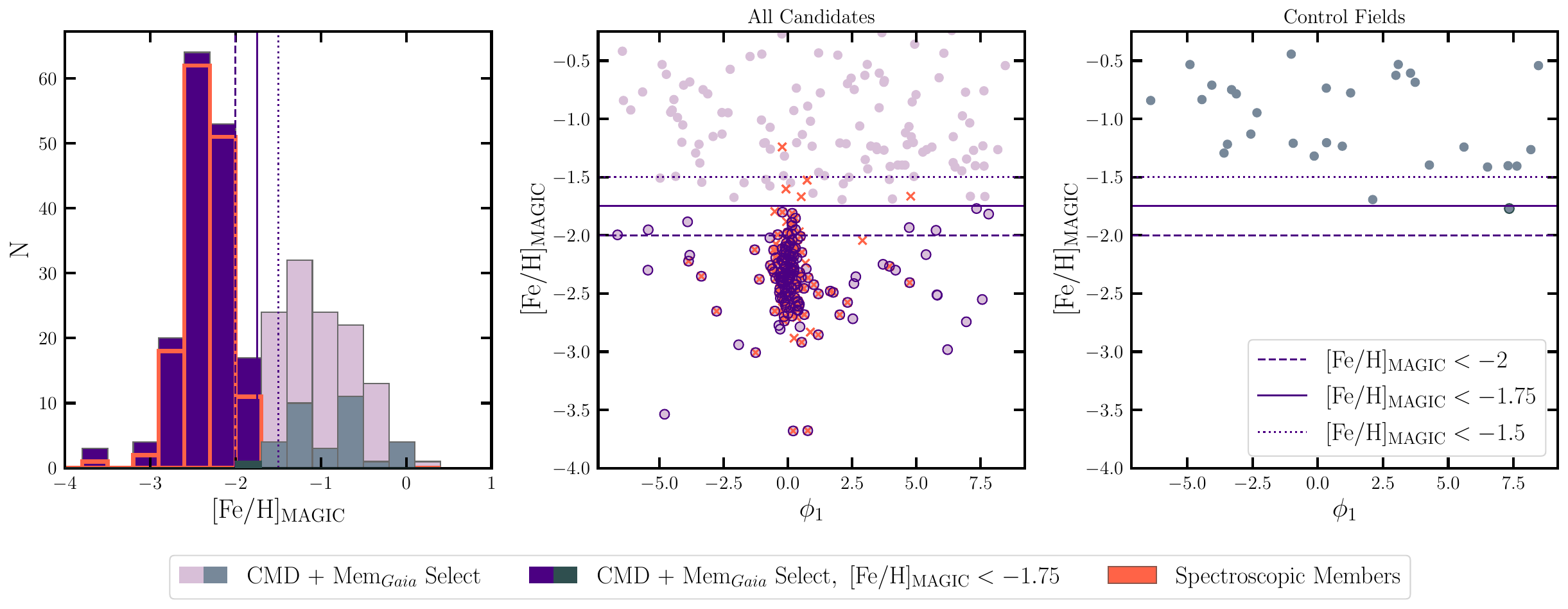}
\caption{Comparison of MAGIC CaHK photometric metallicity distribution for different selection criteria across all candidates and in the isolated candidate control fields. 
\textbf{Left:} Metallicity distribution function of color-magnitude and membership score selected CraII candidates overall/in the background before (light purple/gray) and after (dark purple/gray) metallicity cuts of [Fe/H] \textless~$-2$ (dashed line), [Fe/H] \textless~$-1.75$ (solid line), and [Fe/H] \textless~$-1.5$ (dotted line) compared to the CaHK metallicities of known spectroscopic members \citep{Ji2021,Limberg25} in orange.
\textbf{Center:} Metallicity across the length of the stream for all CraII candidates remaining before and after the metallicity selection.
\textbf{Right:} Metallicity across the length of the stream for CraII candidates in the 11 candidate control fields remaining before and after the metallicity selection.}
\label{fig:mdf} 
\end{figure*}

First, the CraII MAGIC catalog is cross-matched with \gaia EDR3 \citep{Gaia21} astrometry, providing sufficiently reliable star/galaxy separation, and cleaned by a number of cuts.
We use \gaia astrometric quality cuts \citep{Gaia21} to select well-behaved sources from renormalized unit weight error (\texttt{ruwe}~\textless~1.4; \citealt{Lindegren21}) and sources without significant excess noise (\texttt{astrometric\_excess\_noise\_sig}~\textless~2).
Sources with DELVE magnitudes that are determined to be saturated by \texttt{SExtractor} flags \citep{Bertin96} in the $g_0$, $r_0$, and $i_0$ bands and photometry outside the range of the grid of synthetic photometry used in the photometric metallicity calculation are excluded (\texttt{flags\_g}, \texttt{flags\_r}, \texttt{flags\_i} $< 4$ and \texttt{feh\_extrapolation\_flag} $= 0$).
Candidates are also limited to those that are marked as likely stars by the spread-model-based morphology class (\texttt{extended\_class\_g} $< 2$), have broadband colors ($g_0-r_0$ and $r_0-i_0$) consistent with the stellar locus (\texttt{broadband\_valid} $=$ True), and do not have entries in the \gaia DR3 variable star catalog (\texttt{gaia\_var\_flag} $=$ False; \citealt{Eyer23}).
Finally, they are required to satisfy our detection limits of PSF magnitude in the $g$ band $\leq20.5$ and [Fe/H]$_{\rm MAGIC}$ \textgreater~$-4$.

We select stars that lie along the color-magnitude diagram (CMD) of CraII.
We select a 12.5 Gyr Dartmouth isochrone \citep{Dotter08} with [Fe/H] $= -2.24$ that closely matches the age and metallicity of the majority of CraII stars \citep[][]{Walker19,Ji2021}.
Stars are then selected if they have colors ($m-n$) and magnitudes ($m$) that satisfy the following criterion:
\begin{equation}
  \text{min}\left(
  \sqrt{%
    \begin{array}{l}
      <m-m_{\rm iso}>^2+ \\
      {} <(m-n)-(m_{\rm iso}-n_{\rm iso})>^2
    \end{array}%
  } \right) < 0.1
\end{equation}
for discrete magnitudes along the isochrone m$_{\text{iso}}$ and color $m_{\text{iso}} -n_{\text{iso}}$.
We select stars that pass both $g_0$ vs. $g_0-r_0$ and $r_0$ vs. $r_0-i_0$ color-magnitude cuts using these criteria and broadband $g_0$, $r_0$, and $i_0$ imaging from DELVE. 
For our very small magnitude errors, we approximate 0.1 mag as the maximum acceptable deviation from the Dartmouth isochrone to account for spread in age and metallicity.
CraII contains stellar populations of ages in the range of $10.5-12.5$ Gyr  \citep{Walker19} and this age range combined with the CraII metallicity spread has minimal impact in the RGB.
Figure \ref{fig:cmds} shows the significant narrowing of remaining candidates (purple) in color-magnitude and color-color spaces after these selections are imposed.
The left and center panels of Figure~\ref{fig:cmds} show how the data were chosen around the isochrone. 
In the right panel of Figure~\ref{fig:cmds}, we further illustrate the candidates in CaHK color-space (e.g., ${\rm CaHK} - g - 0.9\times(g-i)~{\rm vs.}~g-i$), where stars at different metallicities separate from each other.
Synthetic photometry from Turbospectrum are drawn as black dot-dashed lines to identify these regions of separate metallicity for fixed $\log{\rm g}=2$ (see Section \ref{sec:N395} for details).
This is a typical surface gravity for our sample, but actual data will differ because $\log{\rm g}$ varies per star.
Our selection encompasses most of the spectroscopic members\footnote{We note that we have identified several bluer stars that were considered members in \citet{Ji2021} and \citet{Limberg25} due to their exclusive use of \gaia photometry.  With the more precise DECam photometry, these stars are more likely Milky Way foreground stars and we have excluded them as members here. These stars have Gaia DR3 source\_id = 3543877571488293248, 3543891624621786368, 3571919447324011520, 3543035487317636352, 3544510615308138496, 3543942477034234752.}.

Next, we apply the membership score\footnote{\url{https://github.com/agabrown/sculptor-dwarf-galaxy-gaiadr3}}, ${\rm Mem}_{Gaia}$, selection from \citet{Tolstoy23} for \gaia proper motions and parallax.
The membership score selects stars that are consistent with the proper motion of CraII and have a similarly unresolved parallax.
${\rm Mem}_{Gaia}$ is calculated as:
\begin{equation}
    {\rm Mem}_{Gaia} = v'(C+D)^{-1}v
\end{equation}
for matrices:
\begin{equation}
    v = \begin{bmatrix}
        \varpi-\overline{\varpi} \\
        \mu_{\alpha\star}-\overline{\mu}_{\alpha\star} \\
        \mu_\delta - \overline{\mu}_\delta
    \end{bmatrix},
\end{equation}
\begin{equation}
    C = \begin{bmatrix}
            \epsilon_\varpi^2 & \epsilon_\varpi \epsilon_{\mu_{\alpha\star}} \xi_{\varpi{\mu_{\alpha\star}}} & \epsilon_\varpi \epsilon_{\mu_\delta} \xi_{\varpi{\mu_\delta}} \\
            \epsilon_\varpi \epsilon_{\mu_{\alpha\star}} \xi_{\varpi{\mu_{\alpha\star}}} & \epsilon_{\mu_{\alpha\star}}^2 & \epsilon_{\mu_{\alpha\star}} \epsilon_{\mu_\delta} \xi_{{\mu_{\alpha\star}}{\mu_\delta}} \\
            \epsilon_\varpi \epsilon_{\mu_\delta} \xi_{\varpi{\mu_\delta}} & \epsilon_{\mu_{\alpha\star}} \epsilon_{\mu_\delta} \xi_{{\mu_{\alpha\star}}{\mu_\delta}} & \epsilon_{\mu_\delta}^2
    \end{bmatrix},
\end{equation}
\begin{equation}
    \text{and}~D = \begin{bmatrix}
        \sigma_\varpi^2 & 0 & 0 \\
        0 & \sigma_{\mu_{\alpha\star}}^2 & 0 \\
        0 & 0 & \sigma_{\mu_\delta}^2
    \end{bmatrix}
\end{equation}
where $\varpi$, $\mu_{\alpha\star}$\footnote{$\mu_{\alpha\star} = \mu_\alpha\text{cos}(\delta)$}, and $\mu_\delta$ are parallax, proper motion in the direction of right ascension, and proper motion in the direction of declination, respectively; $\epsilon$ is their errors; $\xi$ is the correlation between uncertainties; and $\sigma$ is their dispersion.
As shown in Figure \ref{fig:z}, stars are kept if their membership score values are below the upper limit for a $\chi^2$ distribution with three degrees of freedom corresponding to a $3\sigma$ limit for a 1D normal distribution, ${\rm Mem}_{Gaia} = 14.2$.
This reduces the spread in parallax and proper motions to converge toward CraII-like objects.
In Figure \ref{fig:cmds}, we observe fewer of the brightest CraII candidates in color-magnitude space and the reddest candidates in color-color space remaining.
Additionally, we test the membership score cut on the stars from the \citet{Ji2021} spectra.
Given that all of their candidates pass the membership score criteria for CraII candidates, we conclude that this technique is well supported as a selection tool in our study.
While the CraII stream may have a proper motion gradient, a $5~{\rm km~s^{-1}~deg^{-1}}$ gradient comparable to the observed velocity gradient from \citet{Limberg25} is $\sim 0.009~{\rm mas~yr^{-1}~deg^{-1}}$ at the distance of CraII and is an order of magnitude smaller than the proper motion errors of CraII candidates.
\citet{Limberg25} included proper motion gradients in their modeling and did not uncover any.

Lastly, we perform a simple selection from the metallicity distribution function (MDF) to exclude remaining metal-rich MW foreground stars from the color-magnitude and membership score selected sample.
The left panel of Figure \ref{fig:mdf} compares the CaHK-based MDFs of the CraII candidates that were selected via CMD and membership score, and the spectroscopic members from \citet{Ji2021}.
We test three upper limits near the peak of the MDF of spectroscopic members to determine the most appropriate threshold for our selection.
The shape of the CraII data below [Fe/H]$_{\rm MAGIC}$ = $-1.75$ appears to preserve the peak of the MDF while still excluding the majority of metal-rich stars more accurately than the [Fe/H]$_{\rm MAGIC}$ = $-1.5$ or $-2$ limits.
It also best retains the dense region of CraII stars along its length in the center panel of Figure \ref{fig:mdf}.
Therefore, we opt to select stars with ${\rm [Fe/H]_{MAGIC}}<-1.75$ as candidate CraII members to maximize the purity of our sample.
This also agrees with the MDF from \citet{Limberg25} that finds approximately no stars with significant membership probabilities at ${\rm [Fe/H]}>-1.75$.
This results in a sample of 161 candidates which is shown in dark purple in Figure \ref{fig:cmds}.
The remaining stars generally narrow in $r-i$ color and strongly agree with the metallicities of known spectroscopic members.

\subsection{Distance Gradient}\label{sec:gradient}

\begin{figure}
    \centering
    \includegraphics[width=1\linewidth]{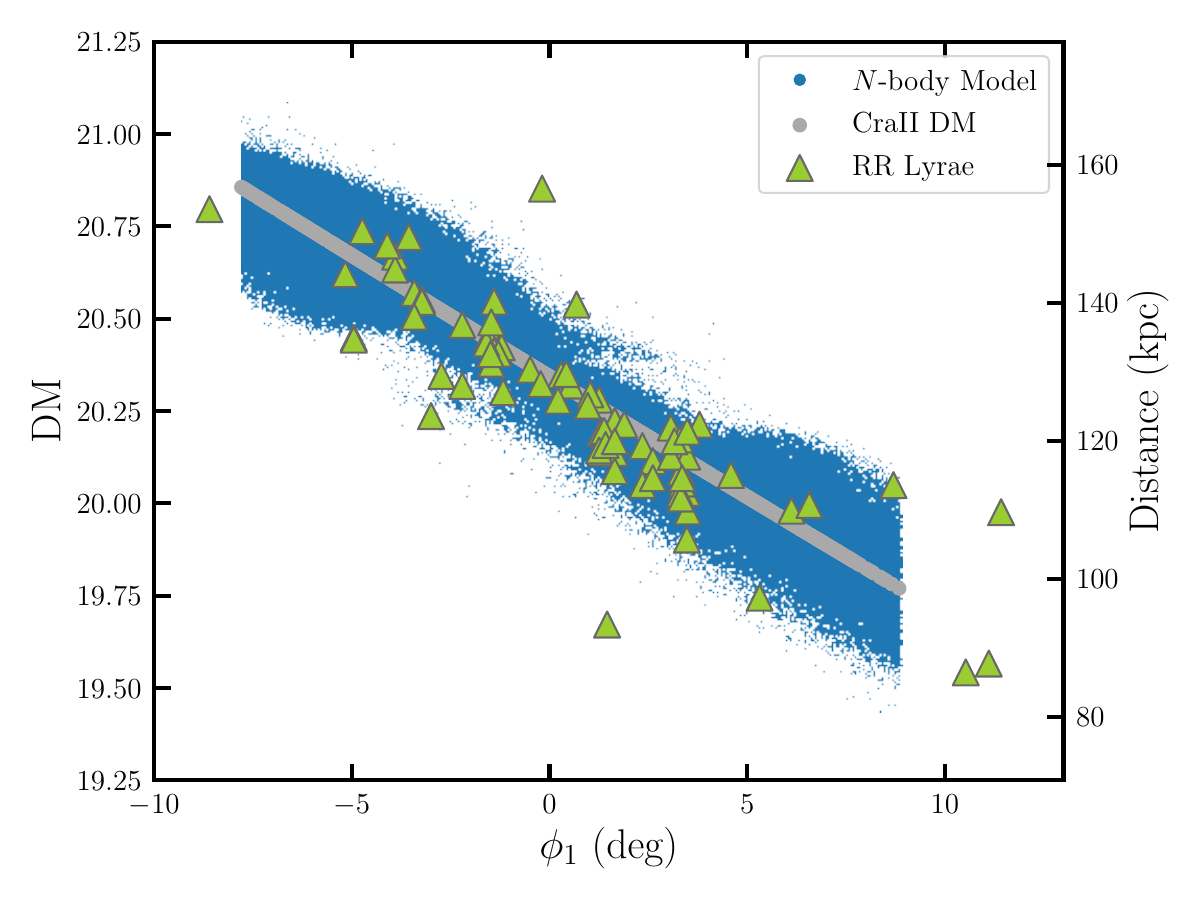}
    \caption{Predicted distance modulus (distance; right axis) as a function of stream longitude from the $N$-body simulation (blue) over the measured area.
    CraII distance modulus from the quadratic fit of $N$-body data across $\phi_1$ is shown in gray for all of the MAGIC quality cut data.
    Candidates from two RRL surveys \citep{Coppi24,Vivas25} are also shown.
    }
    \label{fig:dm}
\end{figure}

\begin{figure*}
\centering
\includegraphics[width=\linewidth]{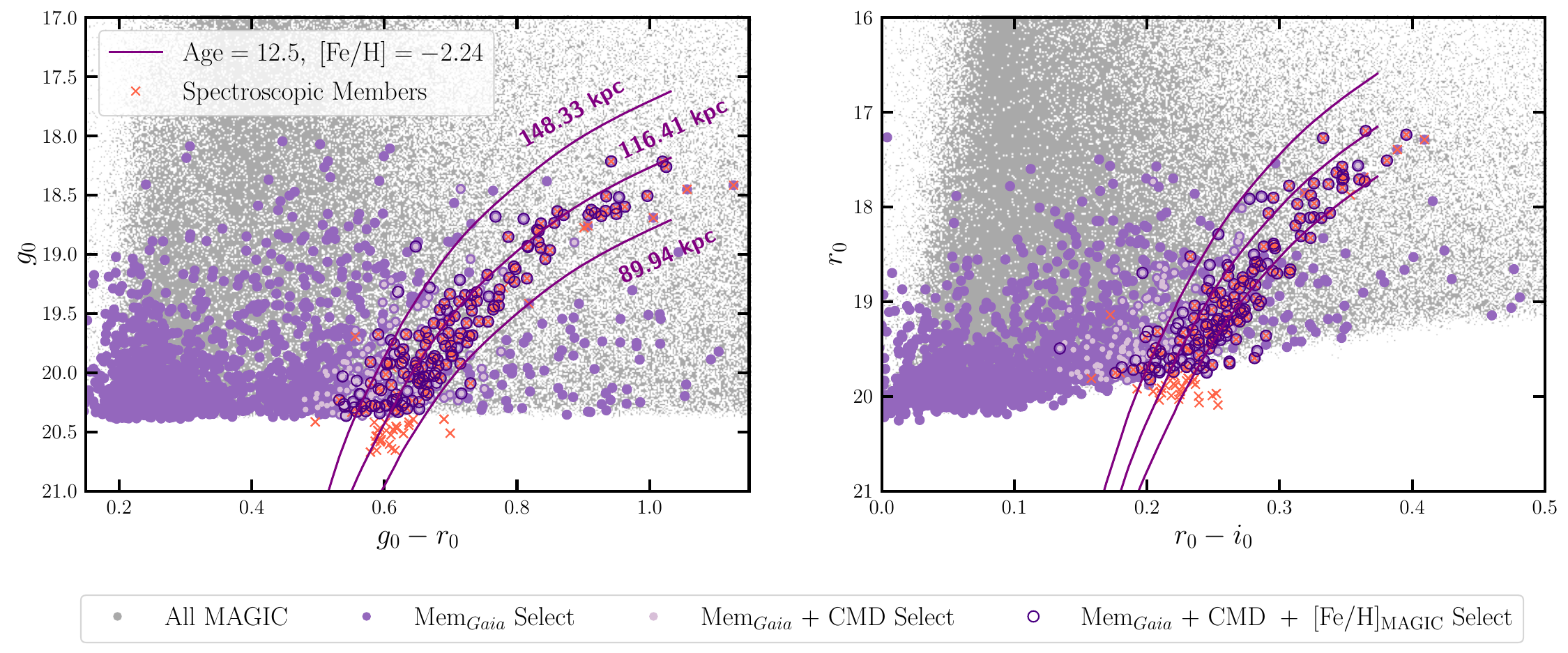}
\caption{Color-magnitude diagrams of DECam $g_0$ and $r_0$ magnitudes against $g_0-r_0$ and $r_0-i_0$ colors.
All quality cut MAGIC data surrounding CraII (as seen in Figure \ref{fig:CraII_pointing}) is shown in gray.
The remaining data after sequential membership score (purple), color-magnitude with a distance gradient (light purple), and metallicity (dark purple) selections are further shown.
The color-magnitude cut selects stars around the isochrones in purple from the Dartmouth Stellar Evolution Database \citep{Dotter08}, considering distance modulus as a function of relative distance from the center of CraII.
Three isochrones are drawn to show the maximum, center, and minimum of the distance modulus range (DM $=19.77-20.86$).
Known spectroscopic members \citep{Ji2021,Limberg25} are marked as orange Xs.
Including a distance gradient retains a wider selection of candidates in CMD space than the selection in Figure \ref{fig:cmds} and better agrees with the known spectroscopic members.}
\label{fig:cmd_grad}
\end{figure*}

In the previous section, our selection required that CMD selected candidates must be consistent with an isochrone at a distance modulus of 20.33 \citep{Torrealba16,Vivas20,Coppi24,Pace25a}, which corresponds to the center of CraII.
However, this does not account for the heliocentric distance gradient along the CraII tidal tails, for which a measurement exists from the \citet{Coppi24} and \citet{Vivas25} studies of RRL.
Specifically, \citet{Coppi24} find a 67.3 kpc difference between the nearest and farthest RRL star. 
In our $N$-body model, we identify a similar correlation along the stream length and with distance modulus.
Figure \ref{fig:dm} compares this $N$-body model to the \citet{Coppi24} and \citet{Vivas25} RRL stars and shows they exhibit a consistent distance gradient.
As discussed in Section \ref{sec:selections}, there should also be small changes in proper motion along the tails, but we find that including these as a function of $\phi_1$ does not alter our selection within this footprint and the differences are consistent with 0 in the \citet{Limberg25} data.

We update the selection in Section \ref{sec:selections} to account for the distance gradient along the stream as follows.
Instead of adding a fixed distance modulus to the isochrone in the CMD selection, we adjust the isochrone along the predicted stream based on the distance modulus from the function:
\begin{equation}
{\rm DM}(\phi_1) = -2.76\times10^{-5}~\phi_1^2~ -0.07~\phi_1 + 20.36
\end{equation}
fit to the data points in the $N$-body model shown in Figure \ref{fig:dm}.
However, the distance needs to be defined before selections so that it is the same for both CMD and membership score cuts\footnote{In practice, this only has a small effect because the parallax is always going to be unresolved at the distance of CraII. The difference is between $\sim \frac{1}{160}$ kpc$^{-1}$ and $\frac{1}{80}$ kpc$^{-1}$ from the far to near side.}.
The order is switched so that the membership score selection is first, then CMD criteria are applied with the distance gradient, and finally we make [Fe/H] cuts.
We note that without the distance gradient, the results of the selection are independent of the order they were performed.
The outcome of this updated method is shown in Figure \ref{fig:cmd_grad} for both $g_0$ and $r_0$ magnitudes.
Candidates after each step appear in CMD space where 162 stars are final selected candidates.
The three purple tracks show the CraII isochrone when we use the minimum, central, and maximum distance modulus from the $N$-body distance gradient function in Figure \ref{fig:dm} and labels their physical distances.
We predict the distance gradient from the model to be $-3.56 {\rm ~kpc~deg}^{-1}$ and find that it closely compares to the absolute value of the gradient from \citet{Vivas25}, $3.7 {\rm ~kpc~deg}^{-1}$.
This results in a wider selection than in Figure \ref{fig:cmds} and stronger agreement with some of the outlying spectroscopic members which may reach the nearest end of the tidal tails.

\subsection{Milky Way Foreground Estimation}\label{sec:mw_foreground}

\begin{figure*}
\centering
\includegraphics[width=\linewidth]{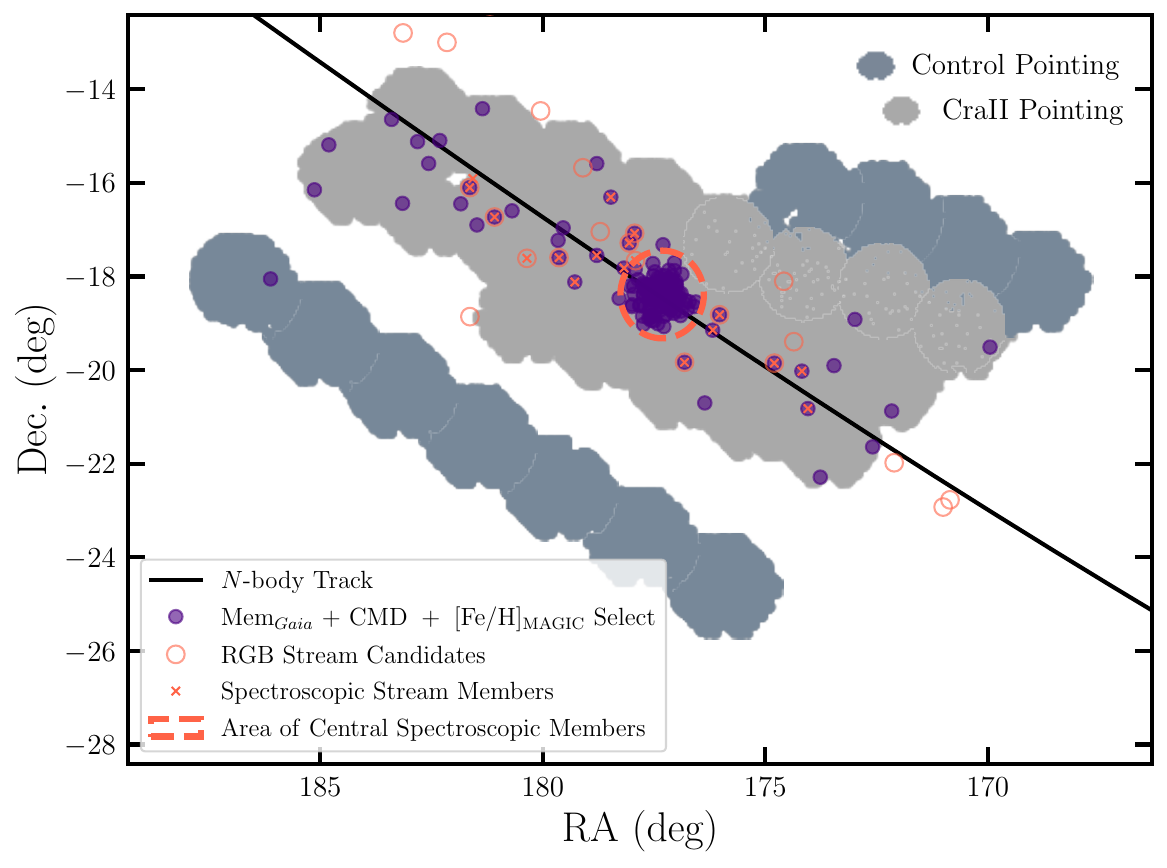}
\caption{Spatial distribution of selected CraII candidates.
All quality cut MAGIC data surrounding CraII (as seen in Figure \ref{fig:CraII_pointing}) is shown in gray.
The remaining data after sequential membership score, color-magnitude with a distance gradient, and metallicity selections is further shown in dark purple.
RGB stream candidates from \citet{Coppi24} are marked as open orange circles.
Known spectroscopic members in the center are circled in orange \citep{Ji2021} and in the stream are marked as orange Xs \citep{Limberg25}.}
\label{fig:cra2_met_cut_mems}
\end{figure*}

\begin{figure*}
\centering
\includegraphics[width=\linewidth]{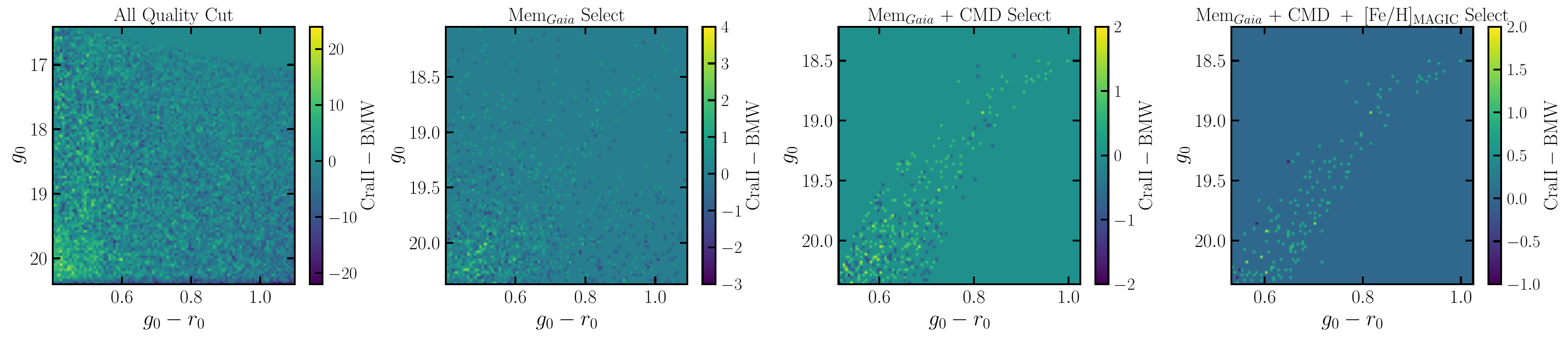}
\caption{Color-magnitude diagram with DECam $g_0$, $r_0$, $i_0$ photometry comparing the number of CraII candidates to MW foreground.
We apply the same BMW limits of $16 \leq r_0 \leq 22$, $-0.25 \leq r_0-i_0 \leq 0.75$, and $0.40 \leq g_0-r_0 \leq 1.10$ to the CraII data.
In each subfigure, all the quality cut MAGIC data surrounding CraII (as seen in Figure \ref{fig:CraII_pointing}) and BMW data are shown in the first panel.
The remaining CraII/BMW data after sequential membership score (second panel), color-magnitude with a distance gradient (third panel), and metallicity (fourth panel) selections are further shown.
The color scale represents the number of remaining CraII candidates subtracted by BMW candidates.}
\label{fig:bmw}
\end{figure*}

\gaia astrometry filters out a large number of MW interlopers and our selection indicates that the inclusion of CaHK metallicity reduces them by an additional $\sim 60\%$.
This is further benefited from the fact that CraII is far from the Galactic plane ($b \sim 34^{\circ}-48^{\circ}$).
However, we are interested in as pure of a CraII sample as possible to measure the width of the stream tails.
Here, we evaluate three ways of estimating the purity of the CraII selection.
We use the number of candidates appearing in our control pointings to calculate the foreground contamination in our CraII fields, compare our fraction of CraII candidates to the fraction of candidates in the \citet{Ji2021} and \citet{Limberg25} spectroscopic studies, and simulate MW contamination from the Besan\c{c}on Model of the Galaxy \citep{Robin86,Czekaj14,Robin17}.

\subsubsection{Empirical Background}\label{sec:emp_background}

We can estimate the foreground contamination in our sample of CraII candidates by testing how many stars pass our selection in the 11 control fields.
In Figure~\ref{fig:cra2_met_cut_mems}, the line of 7 most distant control pointings are at a fixed distance $\sim 3^\circ$ from the system in stream track coordinates (see Section \ref{sec:stream_track}).
There are an additional 4 upper pointings outside of the footprint.
We apply the same CraII membership selection described in Sections~\ref{sec:selections} and \ref{sec:gradient} to the control pointings, which span $\sim 35$ deg$^2$.
The $N$-body model predicts no candidates in this region, and we find 1 star consistent with CraII membership across all control fields.
Assuming this background fraction is constant throughout our CraII pointings (a larger area of $\sim 63$ deg$^2$), we expect just $\sim$ 2 of our 162 CraII candidates may be foreground/background MW stars.
Here we also assume that each pointing is approximately circular with a radius of 1$^\circ$ to determine these areas.
We consider that 6 control pointings in the background and 1 of the 3 CraII pointings outside of the highest quality data outline in Figure \ref{fig:CraII_pointing} are not dithered.
The remaining 25 dithered pointings are 1 magnitude deeper than these 7. 
Therefore, we note that a small number of additional faint stars may be detected if all of the pointings were dithered equivalently, but we do not expect this to significantly change our results.
In Figure~\ref{fig:cra2_met_cut_mems}, our CraII candidates that meet all of the selection criteria are drawn in dark purple.
The number of candidates that are unlikely members is only 2 of 162. 
This gives a low contamination fraction of 1.2\% and is similar to the purity in CaHK-selected members found in \citet{Ji2021} (1 contaminant of 162).
Especially notable are the tidal tail candidates on the far side in Figure \ref{fig:cra2_met_cut_mems} (lower leftmost tail) where no candidates are likely in the background from our selections outside of the footprint in the nearby area.

\subsubsection{Comparison to Spectroscopic Studies}\label{sec:spectra}

Another measure of MW foreground and our contamination rate is available by comparing to CraII spectroscopy \citep{Ji2021, Limberg25}.
\citet{Ji2021} compute membership based on mixture models including radial velocities, spectroscopic metallicities, and \gaia DR3 proper motions, with color-magnitude and \gaia DR3 parallax selections.
From 207 stars with good quality measurements in the center of CraII, they identify 141 members.
The cross match of the \citet{Ji2021} spectroscopic sample and our CaHK sample matches 204 of the 207 stars in the spectroscopic sample and the three missing stars are all spectroscopic non-members. 
After applying our candidate selection, we are left with 109 candidates. 
The missing spectroscopic members are primarily excluded due to the magnitude limit in our CaHK selection. 
Of the 109 cross-matched candidates, two stars are spectroscopic non-members\footnote{Gaia source\_id = 3543812837741109376, 3543818408315571968} implying a purity rate of $98^{+2}_{-5}\%$ at 95\% confidence with our CaHK selection using a Wilson score confidence interval. 
We note that one of the non-members, 3543818408315571968, could be a CraII member if it is an unresolved binary star as its radial velocity is within $15~{\rm km~s^{-1}}$ of the mean velocity of CraII and its spectroscopic and photometric metallicity is consistent with CraII but is a $5\sigma$ outlier due to the low velocity dispersion of CraII. 
However, there are no other velocity measurements of this star \citep{Caldwell17, Fu19, Walker2023ApJS..268...19W, Limberg25} and it should be considered a non-member with current data.

\citet{Limberg25} compute membership with mixture models using radial velocities, spectroscopic metallicities, and \gaia DR3 proper motions. 
Notably, their mixture model includes the tidal tails where stream density is a lot lower compared to the member density in the central field analyzed by \citet{Ji2021}.
Their metallicities are also derived with spectroscopic templates using \texttt{rvspecfit} \citep{Koposov19b} whereas \citet{Ji2021} used Calcium Triplet equivalent width measurements to derive the spectroscopic metallicity. 
\citet{Limberg25} observed 1169 stars with successful velocity measurements and 1010 of these stars cross match with our CaHK sample. 
In contrast to the \citet{Ji2021} sample, our CaHK footprint does not cover the entire spectroscopic sample and part of the missing stars are due to the different footprints. 
Overall, \citet{Limberg25} identify 143 spectroscopic members and 142 are located in the CaHK cross match. 
Of the 1010 stars in the cross matched sample, 128 pass our CaHK selection. 
Of these, 112 are considered members in the \citet{Limberg25} analysis with membership probability $p>0.5$. 
After examining the 10 spectroscopic non-members (8 core, 2 tails) with intermediate membership probabilities \citep[$0.1 <p<0.5$ in][]{Limberg25}, we find that all 10 are CraII members and their low membership probability is due to their higher metallicity from low S/N spectra ($\sim3-7$) and difficulty modeling the low stream density. 
In addition, the 8 core stars are all considered members in the \citet{Ji2021} analysis.  
Of the 6 remaining spectroscopic non-members ($p<0.1$), 4 are located in the core.  
Two of these core stars correspond to the spectroscopic non-members identified in the \citet{Ji2021} comparison. 
The other two, 3543934643014272384 and 3543968659154612224, are $\approx5\sigma$ and $\approx1\sigma$ velocity outliers, respectively. 
Both have low spectroscopic S/N and higher metallicities. 
One, 3543934643014272384, could be an unresolved binary but should be considered a non-member given current data and the other, 3543968659154612224, would be considered a member if the CaHK metallicity is correct. 
There are two non-members in the tails in the cross match, 3567827649160791936 and 3568917677500430848. 
They have offsets in velocity of $\sim1.5\sigma$ and $\sim3.7\sigma$, respectively, and [Fe/H] $\sim -1.5$. 
With the CaHK metallicity, 3567827649160791936, is more consistent with the properties of the CraII system and the lower membership in the mixture model is due to the higher spectroscopic metallicity.
With 6 non-members in the cross-matched sample, this implies a purity of $95^{+5}_{-3}\%$ at 95\% confidence in our CaHK selection using a Wilson score confidence interval. 
If we only consider stars in the tails ($\lvert \phi_1\rvert >1^\circ$), the purity is $10/12 \sim83\%$.

Overall, we find contamination rates of 2-17\%, depending on the spectroscopic sample we compare to and whether we limit the overlap to the CraII stream.  
With our CaHK selection of 37 CraII stream candidates, 6 of these candidates are estimated to be non-members given a contamination rate of 17\%. 
Interestingly, this matches the number of background stars in our width measurements in Section~\ref{sec:width}.

\subsubsection{Besan\c{c}on Model}\label{sec:bmw}

Finally, the Besan\c{c}on Milky Way\footnote{\url{https://model.obs-besancon.fr/modele_ref.php}}  \citep[BMW;][]{Robin86} is a synthetic model of stars in the MW that can be used to generate mock observations. 
We generate lists of stars from the model within each given tile in the outlined footprint of Figure \ref{fig:CraII_pointing}.
BMW stars are generated assuming an initial mass function, star formation rate along the age of the Galactic disc, and an age-metallicity relation \citep{Robin03}.
This includes positions, photometry, astrometry, kinematics, and intrinsic properties.
BMW photometry allows for one apparent magnitude passband and up to four colors.
To best compare to our data, we use errors from functions of CraII's $g_0$ magnitude in the MAGIC catalog rather than the parabolic photometric error functions or fixed proper motion and parallax errors.
These fits are explained in Appendix Section \ref{sec:appendix}.

We repeat the selections from Sections \ref{sec:selections} and \ref{sec:gradient} with and without a distance gradient for the BMW data.
The CraII and BMW data are both limited to $16 \leq r \leq 22$, $-0.25 \leq r-i \leq 0.75$, and $0.40 \leq g-r \leq 1.10$.
However, as suggested by Figure \ref{fig:met}, potential MAGIC metallicity errors may exist (e.g., from dwarf/giant degeneracy) that influence how they compare to BMW metallicities.
Since the model is not limited by the same observational constraints as our data, we apply a normalization factor to make the BMW star counts comparable to our observed star counts in a similar magnitude window and reveal how many random MW stars are then expected to pass our selections.
Using the cuts from Section \ref{sec:selections} that exclude the distance gradient allows us to follow the normalization process from \citet{Tollerud13} for BMW foreground.
We do not measure velocities, therefore, we use our membership score as the kinematic selection.
In each DECam pointing, we count the number of CraII stars that pass the CMD cut but not the membership score selection ($\zeta$).
Then, we repeat this with the BMW stars.
The normalization fraction is:
\begin{equation}
    n = \frac{\zeta_\text{BMW}}{\zeta_\text{MAGIC}}.
\end{equation}
This fraction indicates that the BMW foreground counts should be reduced by $\sim 28$\% on average.
In practice, we apply specific fractions per pointings.
Figure \ref{fig:bmw} shows the difference in CraII and BMW stars after each cut in the selection.
If we subtract the normalized BMW from the MAGIC results, we still recover significant CraII candidates in the galaxy center and in 10 of the tidal tails pointings.
We find an average of $\sim 1$ non-member star per tile within the footprint, for a total of 21.
This may not be an appropriate comparison for this study because the Besan\c{c}on Model assumes a spherically symmetric halo and we expect the halo background to vary significantly \citep[e.g.,][]{Bell08}.
We also model stars at large distances where the simulation may be less accurate than our study.
CraII counts are preferentially higher on the near side of the predicted stream.
Overall, the empirical background likely provides the most reliable estimate of MW foreground in our sample and will be used in calculating our principal contamination rates, but the Besan\c{c}on Model importantly demonstrates that we still obtain CraII candidates across the tidal tails if larger MW star counts like those modeled were observed.

\section{Results}\label{sec:membership}

\begin{table*}
\caption{Crater II Photometric Candidates}
\label{tab:members}
\centering
\begin{tabular*}{\textwidth}{@{\extracolsep{\fill}} c c c c c c c @{}}
\hline
\shortstack{\rule{0pt}{2.5ex}Gaia EDR3 Source ID\\~} &
\shortstack{\rule{0pt}{2.5ex}RA\\(deg)} &
\shortstack{\rule{0pt}{2.5ex}Dec.\\(deg)} &
\shortstack{\rule{0pt}{2.5ex}Parallax\\(mas)} &
\shortstack{\rule{0pt}{2.5ex}PM RA\\(mas yr$^{-1}$)} &
\shortstack{\rule{0pt}{2.5ex}PM Dec.\\(mas yr$^{-1}$)} &
\shortstack{\rule{0pt}{2.5ex}CaHK$_0$\\(mag)} \\
\hline
3543885409803666432 & 177.313700 & $-18.635579$ & $-0.53$ & $-0.35$ & 0.31 & 21.25 \\
3544003603008486912 & 177.321081 & $-18.237114$ & $-0.11$ & $-0.12$ & $-0.27$ & 20.76 \\
3543885405509072896 & 177.322693 & $-18.632478$ & $-0.31$ & $-0.37$ & $-0.25$ & 20.50 \\
3544311053948096512 & 177.296808 & $-17.321098$ & 0.11 & $-0.01$ & 0.21 & 20.45 \\
\ldots & \ldots & \ldots & \ldots & \ldots & \ldots & \ldots \\
\hline
\end{tabular*}
\begin{tabular*}{\textwidth}{@{\extracolsep{\fill}} c c c c c c c c c c @{}}
\hline
\shortstack{\rule{0pt}{2.5ex}$e_{\rm CaHK}$\\(mag)} &
\shortstack{\rule{0pt}{2.5ex}$g_0$\\(mag)} &
\shortstack{\rule{0pt}{2.5ex}$e_g$\\(mag)} &
\shortstack{\rule{0pt}{2.5ex}$r_0$\\(mag)} &
\shortstack{\rule{0pt}{2.5ex}$e_r$\\(mag)} &
\shortstack{\rule{0pt}{2.5ex}$i_0$\\(mag)} &
\shortstack{\rule{0pt}{2.5ex}$e_i$\\(mag)} &
\shortstack{\rule{0pt}{2.5ex}[Fe/H]$_{\rm MAGIC}$\\(dex)} &
\shortstack{\rule{0pt}{2.5ex}$e_{\rm [Fe/H]}$\\(dex)} &
\shortstack{\rule{0pt}{2.5ex}Mem$_{\rm Gaia}$\\~} \\
\hline
0.03 & 20.32 & 0.01 & 19.75 & 0.01 & 19.58 & 0.00 & $-2.05$ & 0.22 & 1.54 \\
0.03 & 19.42 & 0.00 & 18.69 & 0.01 & 18.39 & 0.00 & $-2.03$ & 0.17 & 1.84 \\
0.02 & 19.32 & 0.00 & 18.58 & 0.01 & 18.30 & 0.00 & $-2.12$ & 0.18 & 5.78 \\
0.01 & 19.68 & 0.00 & 19.09 & 0.01 & 18.84 & 0.00 & $-2.92$ & 0.20 & 3.84 \\
\ldots & \ldots & \ldots & \ldots & \ldots & \ldots & \ldots & \ldots & \ldots & \ldots \\
\hline
\end{tabular*}

\vspace{0.5em}

\parbox{\textwidth}{\footnotesize
This table is available in its entirety in machine-readable form at the following Zenodo repository: \url{https://doi.org/10.5281/zenodo.16804863}.  
Candidates are listed in order by relative distance from the center of CraII.  
PMs are the proper motions in right ascension and declination; CaHK$_0$ is the de-reddened magnitude from the N395 filter; $g_0$, $r_0$, and $i_0$ are the DELVE broadband de-reddened magnitudes; [Fe/H]$_{\rm CaHK}$ is the photometric metallicity derived from the CaHK filter; $e_{\rm CaHK}$, $e_g$, $e_r$, $e_i$, $e_{\rm [Fe/H]}$ are their errors; and Mem$_{\rm Gaia}$ is the membership score from Section~\ref{sec:selections}.
}
\vspace{0.5em}
\end{table*}

We apply our methodology to the CaHK dataset and identify 162 CraII candidates out of a total of 182641 initial stars.  
168037 were removed by the initial CMD selection, 14323 were removed by the membership score cut, and 120 were removed by the metallicity selection. 
This provides an overall foreground rejection power of $\sim 10^4$ where $\sim 92\%$ of stars are removed by the CMD cut, $\sim 98\%$ of the remaining stars are removed by the membership score cut, and $\sim 45\%$ of the stars remaining after both other selections are removed by the metallicity cut.
We present these CraII candidates in Table \ref{tab:members} which provides each star’s position, proper motions, magnitudes and errors, metallicity and error, and membership score. 
The full table is available online\footnote{\url{https://doi.org/10.5281/zenodo.16804863}}.
We identify CraII candidates across our observed footprint, arguing for the existence of active tidal disruption.
In this section, we discuss the results of our selection in the center and tidal tails of the galaxy.
In each DECam pointing we recover a number of candidates where we find that some small percentage could be MW contamination (see Section \ref{sec:mw_foreground}).
From these estimates, we analyze the metallicity distribution and measure the width across candidates in the tidal tails to potentially constrain the density profile of CraII's dark matter halo.

\subsection{Final Crater II Candidate Selection}\label{sec:selected_cands}

The center of CraII within a $\sim 1^\circ$ radius ($\sim 2~r_h$) of the central right ascension and declination from \citet{Torrealba16} contains 124 of the 162 detected candidates.
This appears as a central overdensity in Figure \ref{fig:cra2_met_cut_mems}.
Of this, if a maximum of 2 were MW interlopers as calculated in Section \ref{sec:emp_background}, then the contamination rate is at most 1.6\% but is likely lower as the foreground should be approximately uniform.
Some of our candidates identified in the center are new relative to the \citet{Ji2021} spectra.
Altogether, we uncover 124 CraII stars here, 107 of which were previously identified spectroscopically and 17 of which are new.

Of the remaining candidates, 37 are among the DECam pointings in the predicted tidal tails and 1 is from the control fields on the near side.
The four upper control pointings in Figure \ref{fig:CraII_pointing} may also be considered part of the tidal tails, as they are near the outskirts of our $N$-body model.
However, we find that including them does not change our measurements of the stream width.
If both possible MW interlopers from the empirical background estimate were in the tidal tails instead of the center, then the maximum stream contamination rate would be 5.4\%.
We find many new CraII stream candidates and some overlapping with prior RGB stream candidates \citep{Coppi24} and the $S^5$ spectroscopy \citep{Limberg25} as illustrated by Figure \ref{fig:cra2_met_cut_mems}.
Figure \ref{fig:mdf} compares the metallicities of stars in the control fields to the 37 in the stream, showing that the stream candidates are generally more metal-poor than the background, and the one selected star in the control fields only narrowly passes the metallicity threshold.
Of our stream candidates, $\sim 66\%$ (24) come from the near tail ($0.5^\circ<\phi_1<8^\circ$) and the other $\sim 34\%$ (13) are on the far side ($-8^\circ<\phi_1<-0.5^\circ$).
Although it is a difficult comparison given the smaller sample of background stars, the stream peaks around the CraII mean [Fe/H] $= -2.16$ from \citet{Ji2021} whereas the foreground stars are more randomly distributed. 
We see these stream stars as a diffuse distribution of candidates along the stream track in Figure \ref{fig:cra2_met_cut_mems}.
Given the central stellar mass estimate of $\sim 10^{5.55} ~{\rm M}_\odot$ from \citet{Ji2021} for the central system, where we measure 124 stars, we predict a lower limit of stellar mass in the tails of $1.06\times10^5~{\rm M}_\odot$ from the fraction of CraII candidates that we see beyond $\sim 2~r_h$ of the system.
This suggests CraII has lost $\gtrsim 25\%$ of its initial stellar mass.

\subsection{Metallicity Distribution Function of Crater II}

\begin{figure}
    \centering
    \includegraphics[width=\linewidth]{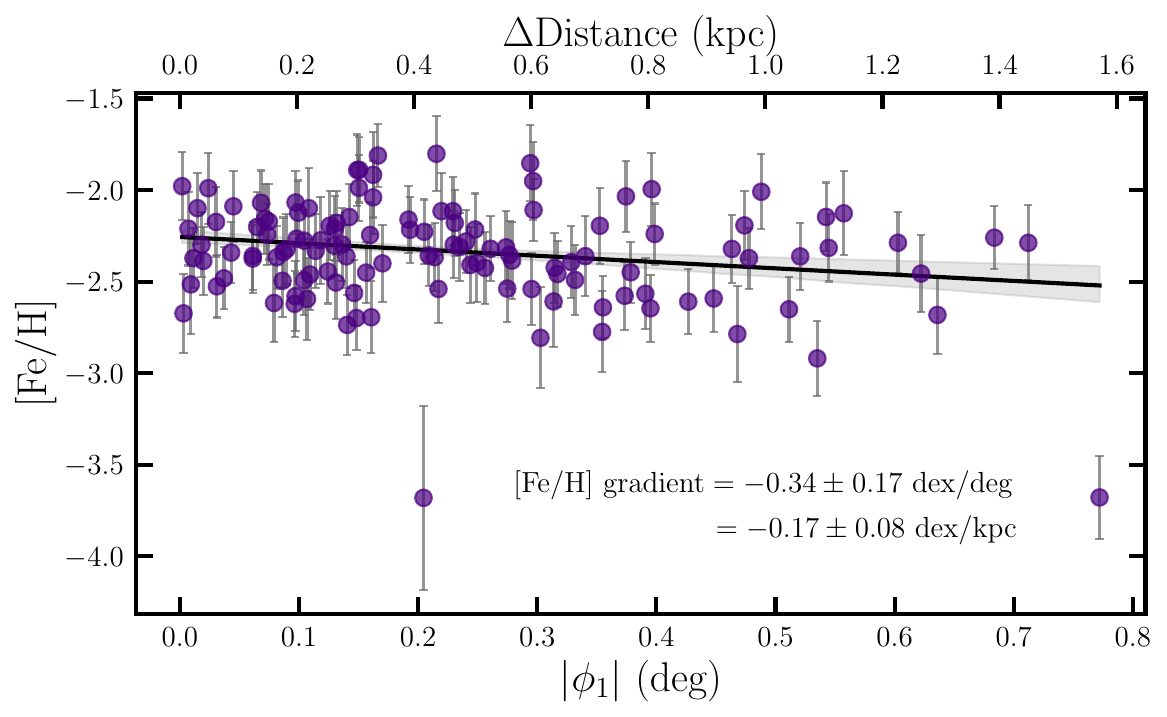}
    \caption{Metallicity gradient (black line) of CraII candidates (purple) along $\phi_1$ in the center pointing of the galaxy, $-0.34\pm0.17~{\rm dex}~{\rm deg}^{-1}$ ($-0.17\pm0.08 ~{\rm dex}~{\rm kpc}^{-1}$).
    A shaded region indicates the $1\sigma$ error on the gradient.
    There are two outliers with [Fe/H] $< -3.5$.
    Removing them would flatten the gradient by $\sim 0.1$ dex, but it would still be greater than a $1\sigma$ detetction. 
    However, we believe they should not be excluded given their errors and the sensitivity of the N395 filter to low metallicities.}
    \label{fig:met_gradient}
\end{figure}

We present the MDF of all of our CraII candidates in Figure \ref{fig:mdf}.
We find an average ${\rm [Fe/H]}_{\rm MAGIC} = -2.37 \pm 0.02$ and a metallicity dispersion of $\sigma_{\rm [Fe/H]}=0.32\pm0.02$ from the standard deviation of CraII candidates' CaHK metallicities.
We determine that the metallicity gradient is not statistically significant ($0.00\pm0.01 ~{\rm dex}~{\rm deg}^{-1}$) using a weighted linear regression analysis of metallicity as a function of $\phi_1$ distance with bootstrapped error estimates.
We compare to literature values in the center of CraII.
\citet{Caldwell17} obtain [Fe/H] $= -1.97\pm0.1$, $\sigma_{\rm [Fe/H]}=0.22^{+0.04}_{-0.03}$, and no evidence of a metallicity gradient; \citet{Fu19} determine [Fe/H] $= -1.95^{+0.06}_{-0.05}$ and $\sigma_{\rm [Fe/H]}=0.18^{+0.06}_{-0.08}$; \citet{Vivas20} find $\sigma_{\rm [Fe/H]}=0.17$; and \citet{Ji2021} measure [Fe/H] $= -2.16\pm0.04$, $\sigma_{\rm [Fe/H]}=0.24\pm0.05$, and no metallicity gradient.
\citet{Limberg25} measure [Fe/H] $= -2.16\pm0.04$  and $\sigma_{\rm [Fe/H]}=0.28\pm0.03$ in the CraII core and their model includes metallicity gradient in the CraII stream (core and tails) but does not detect one.
We similarly do not identify a gradient across the full sample, but our results are more metal-poor than any previous studies and have higher metallicity dispersion.
We attribute this difference to the metallicity offset identified in Figure \ref{fig:met} and defer to the spectroscopic metallicity as a more accurate measurement.
However, there may also be a difference from the large extent of our footprint, covering regions of the tidal tails more distant from the center.
Therefore, we additionally consider our data separated into the central CraII dSph and disrupting CraII stream.

In the center ($0.0 \lesssim \phi_1 \lesssim 0.8$), we find an average ${\rm [Fe/H]}_{\rm MAGIC} = -2.34 \pm 0.03$, a metallicity dispersion of $\sigma_{\rm [Fe/H]}=0.28\pm0.02$, and a metallicity gradient of $-0.34\pm0.17 ~{\rm dex}~{\rm deg}^{-1}$.
We perform the same linear regression analysis with bootstrapping to calculate the metallicity gradient.
The metallicity from the intercept of this function at $\phi_1=0^\circ$ is $-2.26 \pm 0.04$, slightly differing from the average.
Although our result is still more metal-poor than previously measured, the mean metallicity might be biased lower because we select stars with a metallicity cut, [Fe/H] \textless $-1.75$.
Given this and the dispersion, it does overlap with prior estimates, however, the precision of our metallicities may reveal that CraII is more metal-poor than previously estimated.
The metallicity gradient also becomes more prominent in this central region.
Figure \ref{fig:met_gradient} shows the metallicities of CraII candidates in our center pointing as a function of $\phi_1$ and the negative gradient we calculate overplotted.
At a distance of $\sim 116.6$ kpc to the center of CraII \citep{Vivas20}, this gradient is $-0.17\pm0.08 ~{\rm dex}~{\rm kpc}^{-1}$.
The detection of a metallicity gradient suggests that CraII's chemical enrichment may have occurred some time after the galaxy was $\sim3$ Gyr old and its RRL stars were formed, since \citet{Vivas20} previously obtained the photometric metallicity distribution of RRL stars in CraII and did not identify a gradient \citep[e.g.,][]{Martinez16}.
There are two candidates approximately 1 dex more metal-poor than the next lowest metallicity star.
Excluding these candidates flattens the gradient by $\sim 0.1$ dex, making it a $\sim 1.4\sigma$ detection.
A limit of [Fe/H] $<-1.5$ flattens the gradient by a similar degree compared to our chosen [Fe/H] $<-1.75$ criteria, but would also still be detectable.
Given their errors and the capability of the N395 filter to detect such low metallicities, we believe these are truly metal-poor stars and would be interesting targets for high-resolution follow-up.

We further find an average ${\rm [Fe/H]}_{\rm MAGIC} = -2.45 \pm 0.06$, a metallicity dispersion of $\sigma_{\rm [Fe/H]}=0.41\pm0.05$, and a metallicity gradient of $0.06\pm0.03 ~{\rm dex}~{\rm deg}^{-1}$ in the stream (excluding the center).
Although this is less notable than in the center and an overall gradient is still mostly undetectable in our footprint, we show that the average stream metallicity is lower than that in the center, suggesting that over a wider area we may begin to observe a metallicity gradient across large distances.
We also make the first strong detection of a metallicity gradient in the center of the galaxy.

\subsection{Crater II Stream Width}\label{sec:width}

\begin{figure*}
\centering
\includegraphics[width=\linewidth]{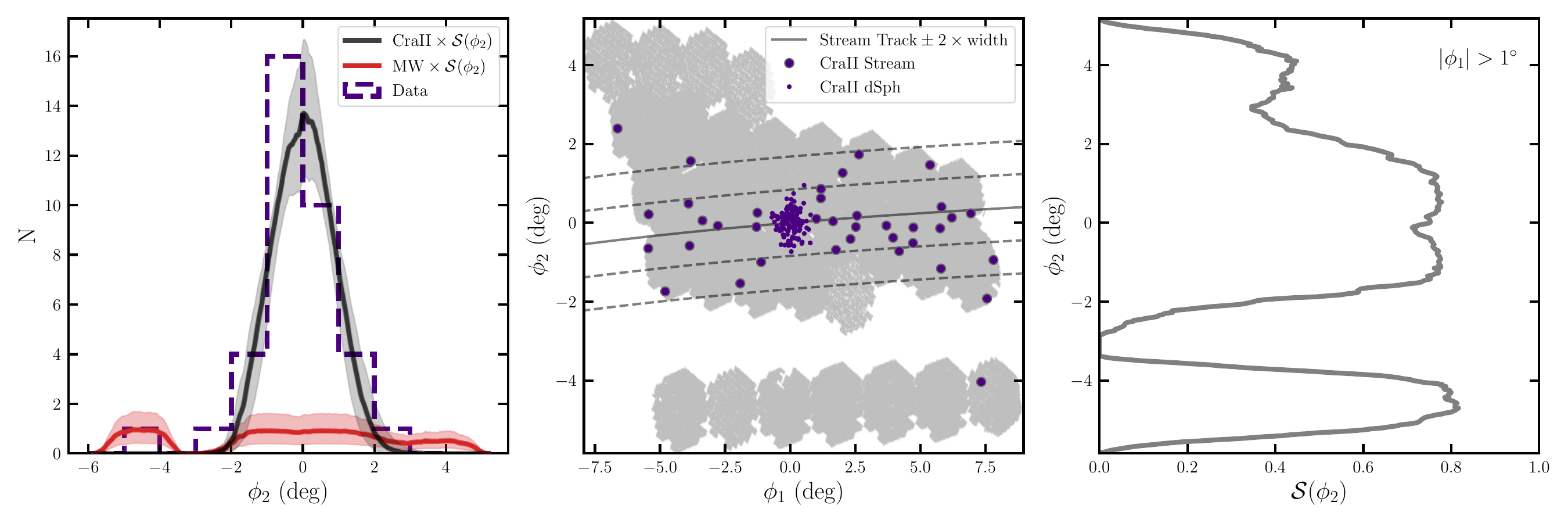}
\caption{Width measurement of the CraII stream from fitting a mixture model that accounts for our non-uniform footprint. 
\textbf{Left:} Histogram of CraII candidates (indigo bins), compared to the CraII model (black line) and the MW foreground model (red). 
Both the CraII and MW models are convolved with the selection function ($\mathcal{S}(\phi_2)$) and the shaded bands denote $1\sigma$ errors. 
\textbf{Center:} Distribution of CraII candidates relative to its stream track and our footprint. 
The dotted lines denote one and two times the median width value. 
\textbf{Right:} Selection function of our footprint. 
We exclude $\lvert \phi_1 \rvert<1^{\circ}$ as it contains the CraII dSph.}
\label{fig:stream_width} 
\end{figure*}

To determine the width of the CraII stream, we fit a mixture model that jointly accounts for the morphology of the stream and the footprint.  
We build a selection function ($\mathcal{S}(\phi_2)$) by building a grid in $\phi_1, \phi_2$ with bin size of 0.025$^\circ$ and a binary option of whether each bin includes stars.
We then sum along the $\phi_1$ axis and normalize by the total number of bins so that the fractional values correspond to the percentage of bins covered. 
We exclude $\lvert \phi_1\rvert<1^\circ$ to remove the CraII dSph. 
The selection function is shown in the right-hand panel of Figure \ref{fig:stream_width}.

The mixture model we fit includes a CraII stream component and MW foreground component.
The  likelihood for star $i$ is given by: 
\begin{equation}
    \mathcal{L} = \mathcal{S}(\phi_2)\left[f~B~\mathcal{N}(\phi_{2,i} - \phi_2^{\rm orbit}(\phi_{1,i}), w) + (1-f)~C \right], 
\end{equation}

\noindent where the first term describes the CraII stream as a normal distribution in $\phi_2$ with an unknown width, $w$, and the second term models the MW foreground as a constant distribution in $\phi_2$. $f$ is the fraction of stars in the CraII stream. We normalize the likelihood such that $\int \mathcal{L}~{\rm d}\phi_2=1$ and $B,C$ are the normalization constants. We account for the slight offset between the orbit and $\phi_2=0^\circ$ by setting the mean of the normal distribution to the orbit. Both $w$ and $f$ are free parameters in this model. 
We fit this model with the Markov Chain Monte Carlo sampler \texttt{emcee} \citep{emcee}.

Applying this model to the 37 candidates with $\lvert \phi_1\rvert>1^\circ$, we find $w=0.80^\circ~\!{}\!{}_{-0.33^\circ(-0.57^\circ)}^{+0.58^\circ(+1.62^\circ)}$; $N_{\rm CraII}=31_{-5}^{+4}$; $N_{\rm MW}=6_{-4}^{+5}$. 
The errors in parentheses give the $2\sigma$ limits on the width.
This corresponds to $w=1.63_{-0.68}^{+1.16}~{\rm kpc}$ if we assume the stream is at the distance of the center of CraII.
With $r_{h}=31.2^{'}$, we find the ratio of stream width to half-light radius to be $\widetilde{w}=1.54_{-0.64}^{+1.10}$. 
We show the model posterior distribution and compare it to the data in the left-hand panel of Figure~\ref{fig:stream_width}.
The central panel compares the candidate stream members, the footprint, and the stream width along the orbit. 
We highlight that candidates are found at all stream longitudes ($\phi_1$), implying that we have not mapped the entire stream and it extends beyond our covered area.
Furthermore, the $N$-body model predicts debris to larger distances.
Given the width we measure, the 4 stars outside of twice the width are expected to be MW foreground stars.
Overall, we find the stream to be wider than CraII and estimate $31/37\sim 83\%$ of the tail candidates to be CraII members, but there are significant uncertainties due to the low surface density of the stellar stream.

\begin{figure*}
\centering
\includegraphics[width=\linewidth]{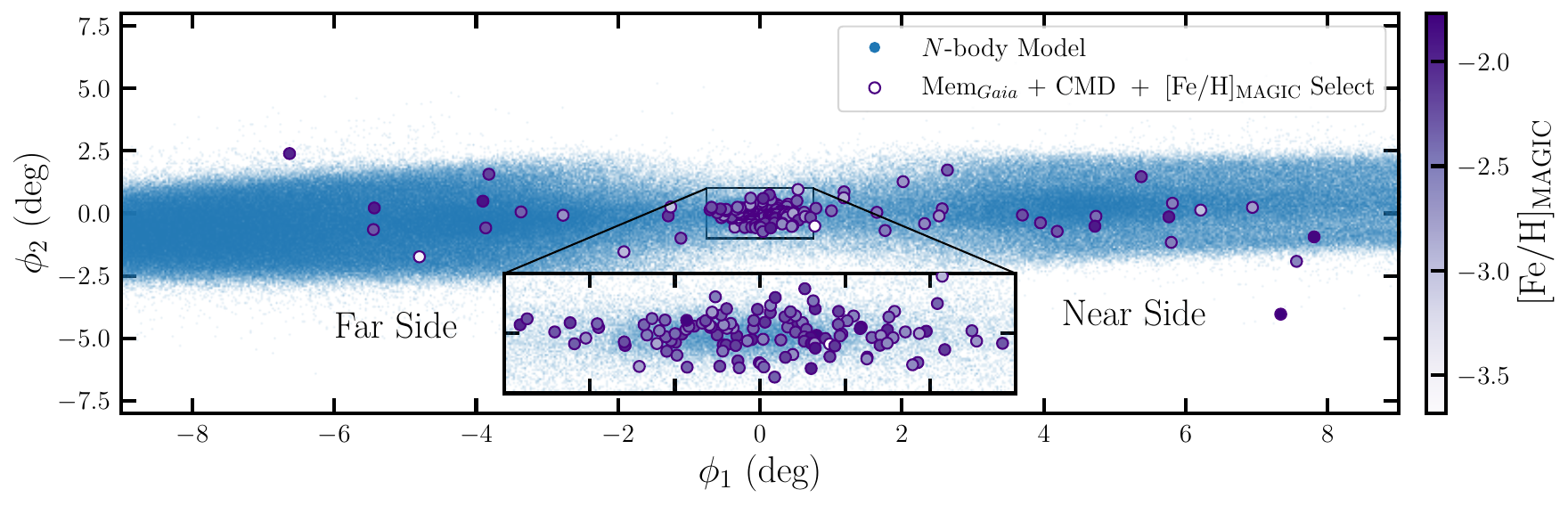}
\caption{Spatial distribution of selected CraII candidates (dark purple outline) compared to the $N$-body simulation (blue) used to choose DECam pointings.
Candidates are color-coded by CaHK metallicity.}
\label{fig:nbody} 
\end{figure*}

We compare the observed stream width to theoretical predictions from the $N$-body model described in Section~\ref{sec:$N$-body}.
We limit the $N$-body model to $-8^\circ$\textless$\phi_1$\textless$8^\circ$, which spans the range of our CraII selected candidates.
The results of the $N$-body model indicate that the system releases nearly double the material on the far side of the tidal tails compared to the near side. 
However, our data finds approximately three times the number of candidates on the near side compared to the far side.
Although it is interesting that we detect the opposite behavior as to what was predicted, the detectability of CraII stars is different on either side of the stream.
The inconsistency could be due to measuring deeper on the near side. 
The near side probes roughly 0.7 mag fainter than the far side. 
While the near side is at higher Galactic latitude ($b\sim43-49^\circ$) compared to the far side ($b\sim34-41^\circ$) and should have lower MW contamination, the CraII stream is overall at high enough Galactic latitude that major differences between the two sides of the stream are not expected. 
The $N$-body model explored here was only `fit' to the properties of the CraII dSph and future $N$-body models can utilize the properties of the stream uncovered here.
The variance of the Gaussian fit to the $\phi_2$ of candidates in the $N$-body model indicates a stream width of $1.19^\circ~\!{}\!{}^{+0.46^\circ}_{-0.71^\circ}$.
The error is determined from minimum and maximum $\phi_2$ across $\phi_1$ bins.
This is approximately 1.5 times the measured width from the final selected photometric candidates, $\sim 0.8^\circ$.
CraII candidates are detected across the full width of the $N$-body model as seen in Figure \ref{fig:nbody}.
The candidates extend from the center to the edges of the observed footprint in length and width.
The one candidate that we find beyond the $N$-body model lies in one of the control pointings.

\section{Discussion}\label{sec:discussion}

Our results detect a significant stellar stream, favoring an active tidal disruption scenario \citep{Sanders18,Fu19,Ji2021,Borukhovetskaya22,Bonaca25}.
Given this, we explore how tidal stripping could occur within different dark matter models to reproduce the observed CraII tails.
Hereafter, we refer to the background observations as the combined 11 control fields (the offset upper 4 and lower 7 tiles shown in Figure \ref{fig:CraII_pointing}) rather than evaluating the 4 upper pointings as a part of the far tail despite their close proximity to the $N$-body model candidates. 
We find that including these 4 upper pointings do not change our results in Section \ref{sec:width}.

\subsection{Tidal Disruption}

We uncover a clear stream across a $16\times8$ deg$^2$ area.
Of our sample of CraII candidates, nearly 25\% of stars are within the tidal tails.
They have been carefully isolated from MW background and neatly trace predictions from $N$-body modeling.
We expect the ratio of RRL to RGB stars to be constant between the center and tidal tails assuming their stars have the same age and metallicity distributions.
We detect candidates in 13 of the same pointings as the RRL stars from \citet{Vivas20}, \citet{Coppi24}, and \citet{Vivas25} which cover some of the furthest measured distances from the center of CraII along its stream track relative to previous work.
The ratio of RGB stars above the horizontal branch from this study to RRL stars from \citet{Vivas20} and \citet{Vivas25} in the center is 0.92:1 and to the \citet{Coppi24} and \citet{Vivas25} RRL stars in the tidal tails is 0.54:1 (0.36:1 on the far side and 0.67:1 on the near side).
We find that these values justify we are observing signal rather than noise or background which would inflate the ratio.
They also indicate that more faint, RGB CraII candidates may remain undetected in the stream because the ratio is higher in the center.
Adding this to what was previously known about CraII's close proximity to the MW at pericenter from proper motion measurements \citep{Sanders18,Fritz18,Kallivayalil18,Fu19,Ji2021,Errani22,Borukhovetskaya22,Pace22}, we confidently claim that CraII is tidally disrupting and present the first photometric metallicity-selected RGB candidates of its underlying stream (see Table \ref{tab:members}).

We use the detection of these stars across the observed CraII footprint to estimate the surface brightness of the stream. 
We estimate the surface brightness of the CraII stream by scaling the CraII flux within the half-light radius by the relative number of stars in the stream compared to the dwarf.  
With our CaHK selection, we identified 97/124 stars in the central pointing within the half-light radius, 13 stars on the far side, and 24 stars on the near side of the stream (i.e. the flux in the stream should be $\approx37/97$ the flux within the half-light radius). 
Accounting for the larger area of approximately 8 and 9 DECam pointings or $\approx 25~{\rm deg}^2$ and $\approx 28~{\rm deg}^2$ on the near and far side, respectively, we estimate $\mu=36.1~{\rm mag~arcsec^{-2}}$ in the overall stellar stream and 35.7 and 36.5$~{\rm mag~arcsec^{-2}}$ on the near and far sides, respectively. 
This estimation assumes that the stellar populations are the same in the stream and core and that the stream and core are probed to the same stellar luminosity.  
This last assumption is invalid given the distance gradient in the stream (i.e. we probe fainter stars in the near side of the stream compared to the far side) and our estimate of the overall stream surface brightness is more accurate than the value on either side.

Several simulations find that low density stellar streams are predicted around most dwarf galaxies \citep{Shipp23}.
The Auriga simulations \citep{Riley25,Shipp25} find that intact satellites are expected to be outnumbered by streams; only surviving if they were recently accreted without experiencing more than one pericenter and have large apocenters ($r_{apo}\gtrsim200$ kpc) and pericenters ($r_{peri}\gtrsim50$ kpc).
Many of these MW dwarf galaxy streams overlap the known population of intact MW satellites in simulations, suggesting that they may be more disrupted than current imaging has revealed.
The remaining bound components of disrupting galaxies are often relatively undisturbed \citep{Wang22} as illustrated in Figure 1 of \citet{Vienneau24}, allowing their diffuse, low surface brightness tidal tails to hide.
In the FIRE-2 \textit{Latte} simulations, \citet{Shipp23} found that disrupting satellites were misclassified due to the low surface brightness of their tidal tails (between 34 and 38 mag arcsec$^{-2}$).
Our confirmation of CraII's low surface brightness stream acts as an example of this idea.
Although its exceptionally unusual properties signaled that tidal stripping or another mechanism was at play, it was previously concealed as a result of being too faint. 
The $\sim 36$ mag arcsec$^{-2}$ tidal tails only became visible by tracing them with RRL stars \citep{Coppi24,Vivas25} and specifically targeting the area with $S^5$ spectroscopy \citep{Limberg25} and metallicity-sensitive photometry.
Other galaxies with even more subtle features would be nearly undetectable using current imaging.
Upcoming deep observations such as those possible with the Rubin Observatory Legacy Survey of Space and Time \citep{LSST09,Ivezic19} and further narrow band surveys may be the key to uncovering more evidence of tidal disruption around existing presumably ``intact" satellites \citep{Wang17,Shipp23}.
Even then, more powerful instruments may be necessary to observationally match the predicted disruption rates.
CraII's faint stream helps bridge this gap between the predicted population of tidal tails around MW satellites and observed MW streams.
The confirmed tidal disruption of this system also opens further discussion about the properties of dark matter that allow for its expected mass loss.

\subsection{Dark Matter Models}

\begin{figure}
    \centering
    \includegraphics[width=\linewidth]{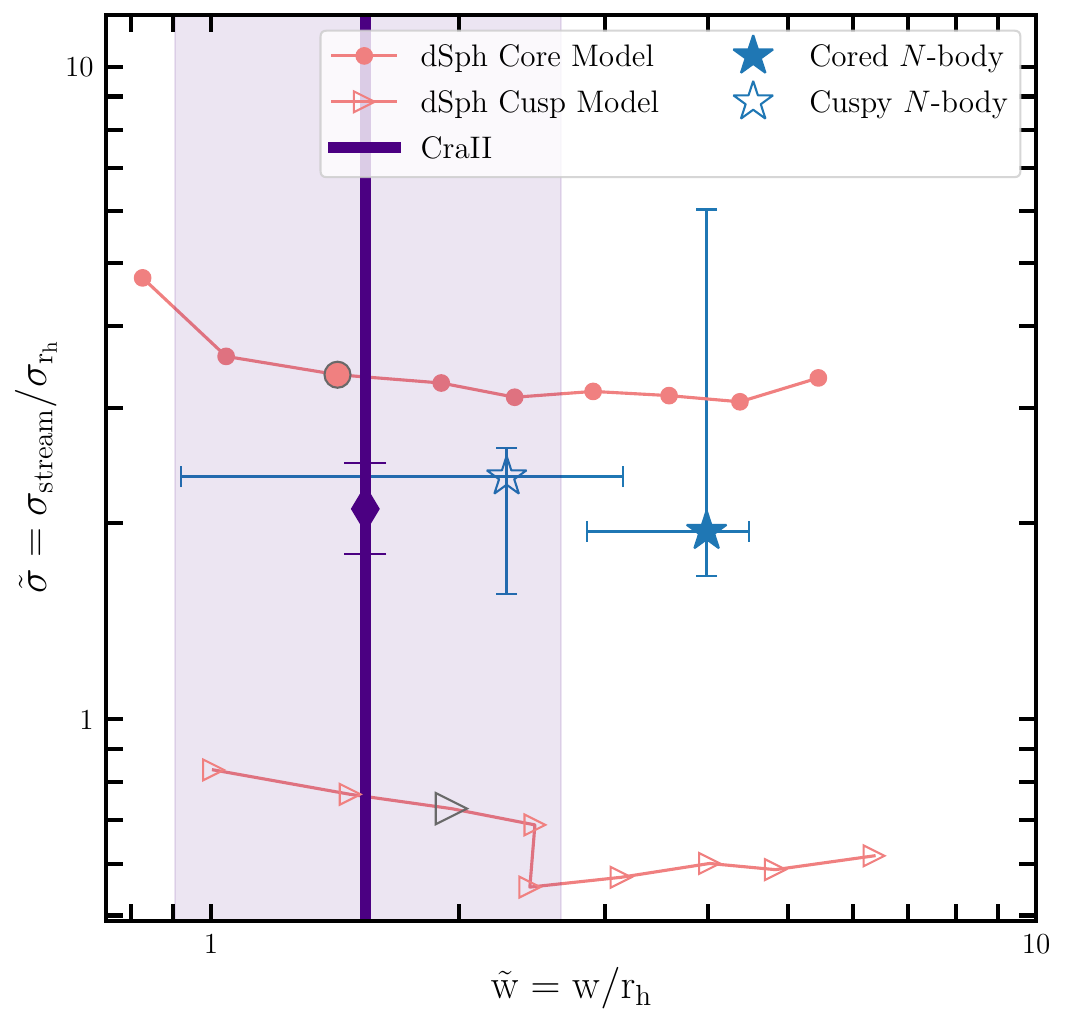}
    \caption{Width and velocity dispersion of the cored (solid circles) and cuspy (open right-facing triangles) dSph models from \citet{Errani15} compared to the CraII width from this paper (dark purple line) and $N$-body modeled (blue) widths.
    The error in CraII width is shown as a highlighted region around the line at the measured value.
    The width is normalized by the half-light radius and the velocity dispersion is a ratio between the stream and center of the galaxy.
    Velocity dispersion from \citet{Limberg25} is marked with their errors as a diamond on the line for our measured CraII width.
    The dSph tracks are modeled at the most recent pericentric passage at which particles become tidally unbound and shown at the fixed snapshot of the 4th apocenter in Figure 2 of \citet{Errani15} from their simulation.
    The enlarged points outlined in gray show the relevant distance for CraII.}
    \label{fig:cusp_core}
\end{figure}

The leading conflict between CraII's large size and low velocity dispersion with $\Lambda$CDM \citep{Sanders18,Borukhovetskaya22} is that the central cusp of the halo should be resilient to tidal stripping.
CraII has such a low velocity dispersion ($\sim 2.5$ km s$^{-1}$) that only strong tides could replicate it, but these would make the galaxy much smaller than observed.
However, if a core were formed from baryonic feedback \citep{Read05,Mashchenko08,Pontzen12,DiCintio14,Chan15}, then these predictions could be modified.
CraII's bound stellar mass approaches the lower limit for cores to form by baryonic processes (e.g., feedback), $\sim 10^6~{\rm M}_\odot$ \citep{Garrison-Kimmel13,Tollet16}.
We compare our new width estimate for CraII with the cored and cuspy tracks of tidal disruption models of dSphs from Figure 3 of \citet{Errani15}.
Their results find that stream width and velocity dispersion relative to what is seen in the central half-light radius of CraII depend on how the dark matter is distributed (i.e., its density profile; see Figure \ref{fig:cusp_core}).
Both the width from our photometric stream candidates and $N$-body model are shown in Figure \ref{fig:cusp_core} where the first is a lower limit until a wider area around CraII is collected to compare it to the maximum from the $N$-body model.
Then, we estimate velocity dispersion for the $N$-body model.
We fit the velocity distribution with a normal distribution with a free mean and dispersion using \texttt{emcee} \citep{emcee}.
The velocity dispersion of the $N$-body model oscillates until settling at distances $|\phi_1|\gtrsim10^\circ$. 
Our error bars represent this as a lower limit of the velocity dispersion range and the upper limit is from the side of our observed tidal tails where CraII candidates have a higher velocity dispersion.
We also repeat this for a cored $N$-body model of CraII (\texttt{core-base} in \citealt{Limberg25}).
The width from this cored model is $\sim 2.07^{+0.26}_{-0.59}~^\circ$, following the procedure described in Section \ref{sec:width}.

The orbits of the dSph models from \citet{Errani15} are not completely equivalent to that of CraII.
They are placed on highly eccentric orbits with a fixed apocenter, $r_{\rm apo} = 150$ kpc, and pericenters $r_{\rm peri} = 5, 7.5, 10, 15,$ and 20 kpc.
\citet{Borukhovetskaya22} finds that CraII's pericenter is poorly constrained with {\it Gaia} DR2 between $\sim 10$ and 50 kpc, last occuring $\sim 1$ Gyr ago.
They explore three orbits at $(r_{\text{peri}},r_{\text{apo}})=(4.23,130)$, (15.5,133), and (37.4,139) kpc from \citet{Kallivayalil18,Fritz18,McConnachie20}, respectively.
The first two orbits used \gaia DR2 proper motions and the third was from \gaia EDR3.
Of these, none could reproduce CraII's surface density profile assuming it intially formed in an NFW halo of virial mass $\text{M}_{200} = 2.7\times10^9~\text{M}_\odot$ unless $r_{\text{peri}} \lesssim 15$ kpc.
\citet{Pace22} and \citet{Battaglia22} additionally measure $(r_{\text{peri}},r_{\text{apo}})= (24.0^{+5.6}_{-5.2},138.1^{+7.9}_{-4.9})$ and (19.9,146.8) kpc, respectively, for CraII including the influence of the LMC.
The snapshot shown from the dSph models uses $(r_{\text{peri}},r_{\text{apo}})=(20,150)$ kpc.

We must assume that CraII's most recent pericenter aligns with the last simulated passage from \citet{Errani15}.
Although their dSph orbit ($r_{\rm peri} = 20$ kpc, $r_{\rm apo} = 150$ kpc) is not exactly the same as CraII's orbit ($r_{\rm peri} = 24.0^{+5.6}_{-5.2}$ kpc, $r_{\rm apo} = 138.1^{+7.9}_{-4.9}$ kpc), it is nearly equal.
Therefore, we compare CraII to the models at the third pericentric passage (light red) in Figure \ref{fig:cusp_core} as this roughly matches the observed period of CraII (2-3 Gyr; \citealt{Battaglia22}).
From left to right, the points represent increasing cylindrical distance to the galactic center between 20 and 100 kpc.
At either end of the measured tidal tails, we reach a distance $\sim 40$ kpc away from the center of CraII.
This would be the third point of the dSph models' third pericentric passage tracks.
Our measured width of the CraII stream better matches their model of a cored dSph than a cuspy dSph.

However, there are significant modeling uncertainties.
\citet{Errani15} model dark matter particles and project the properties of the tidal stream from a probability of tagging stars, whereas we observe CraII's stellar content directly.
They also use a different initial progenitor mass.
We run $N$-body simulations with initial conditions that better represent CraII's properties.
These produce predictions that are inverse to \cite{Errani15}, finding that CraII's width more closely aligns with the cuspy model.
The \citet{Errani15} and $N$-body models may disagree because of large errors, but more modeling is needed to further explore this.
Our data provides important new constraints on these future models, allowing them to provide better boundaries between cuspy and cored halos.

These relationships also illustrate how velocity dispersion may be a key observable needed to differentiate between the two dark matter models.
If it is nearly double the prior measurement ($2.34^{+0.42}_{-0.30}$ km s$^{-1}$) or more, then CraII will be more consistent with the cored halo dSph models.
Less than this, and a cusp would be likely.
\citet{Limberg25} measure the velocity dispersion ratio to be $2.10^\circ~\!{}\!{}^{+0.37^\circ}_{-0.31^\circ} ~{\rm km}~{\rm s}^{-1}$ from spectroscopic members selected with $S^5$.
This is between the velocity dispersion ratios for dSph cusp and core models from \citet{Errani15} and also between the ratios from our $N$-body models.
Binary stars have not been considered in the models or data but would likely inflate the stream velocity dispersion more than the center velocity dispersion, making the overall velocity dispersion ratio lower.
Given the wide error range and disagreement of the models, the effect of binaries would be negligible on our interpretation of CraII's dark matter profile.
We thus determine from this study that CraII's dark matter halo is still ambiguous and could be either cuspy or cored.

Alternative dark matter models can also produce dark matter cores which could help explain CraII's unusual properties.
One alternative model is Self-Interacting Dark Matter (SIDM) and \citet{Zhang24} evaluate the evolution of CraII in the tidal field of the MW in SIDM with $N$-body simulations, modeling the initial halo with an NFW profile \citep{Navarro97}.
They find that with a 1 kpc core and self-interacting cross section of $\sim 60~\text{cm}^2~\text{g}^{-1}$, they could reproduce the observed low velocity dispersion and large half-light radius of the system regardless of its initial stellar distribution.
Comparatively, even if baryonic feedback formed a core in CDM, the FIRE2 simulation predicts that the core size should be much smaller, on the order of 10 pc, for a progenitor halo with a similar mass to CraII's, $3 \times 10^9~\text{M}_\odot$ \citep{Lazar20}.
This would be too small to significantly affect CraII.
Although we are still learning about baryonic properties in dark matter models \citep{Gutcke25,Cruz25}, if these theoretical predictions for CDM vs. SIDM cores are correct, then SIDM may be a viable model of the unusual properties shared by CraII and a growing number of other dSphs.

$\psi$DM, commonly referred to as fuzzy or wave dark matter, has also been previously considered \citep{Schive14,Pozo25}.
\citet{Pozo22} explored $\psi$DM to explain CraII.
Their $\psi$DM model finds a large core of radius $\simeq 0.71^{+0.09}_{-0.08}$ kpc and total mass of $2.93^{+2.99}_{-1.44}~ \times 10^8~\text{M}_\odot$, consistent with the dynamical mass estimate by \citet{Caldwell17}.
Given this, they explain that the extended size of the core is a result of its low halo mass ($\simeq 10^8~\text{M}_\odot$) and significant tidal stripping.
Unlike in CDM \citep{Fattahi18}, tides naturally allow for CraII's low velocity dispersion and large radius simultaneously in $\psi$DM.
A cuspy NFW profile is also inconsistent with the stellar density and kinematic behavior in the central regions of the galaxy as seen in Figure 1 of \citet{Pozo22}.
They find the soliton form combined with a shallow NFW halo to better fit the stellar profile of CraII.
Altogether, they argue that the tidal disruption at CraII's small pericenter follows the expected behavior of $\psi$DM galaxy simulations \citep[e.g.,][]{Schive20}. 
However, a number of constraints on $\psi$DM suggest it is incompatible with existing observations \citep[e.g., Figure 2 of][]{Rogers21,Zimmermann25,May25,Teodori26}.
Ultimately, this unique system is proving to be a useful probe of alternative dark matter models.

\section{Conclusion}\label{sec:conclusion}

We present new DECam CaHK photometry of the CraII dSph and identify CraII candidates within the intact center and along the tidal tails, outlining a near and far aligned stream component predicted by our $N$-body tidal disruption model and prior RRL/spectroscopic observations \citep{Coppi24,Limberg25,Vivas25} and demonstrating the success of the N395 filter in uncovering extremely faint substructures in metal-poor systems like CraII.
We conclude that:
\begin{itemize}
    \item The CraII stellar stream extends fully across the predicted tidal tail region of the $16\times8$ deg$^2$ area in our observed footprint, $\sim 80$ kpc long.
    Since we find candidates throughout our entire footprint, we expect that the stream extends even further along the tidal disruption axes. 
    \item 162 stars pass our color-magnitude, membership score, and metallicity (${\rm[Fe/H]}_{\rm CaHK}<-1.75$) selections  considering the large distant gradient and metallicities from CaHK imaging that allow for the removal of MW foreground.
    Of these, 37 candidates are located in the tidal tails/stream, implying CraII has lost $\gtrsim 25\%$ of its initial stellar mass.
    We measure an extremely faint surface brightness of this stream $\sim 36$ mag arcsec$^{-2}$.
    \item We explore the resulting metallicity distribution functions in the center and stream of CraII. While we do not identify a metallicity gradient in stream, we do find a significant gradient in the main body ($-0.34\pm0.17~{\rm dex}~{\rm deg}^{-1}$).
    \item Our observed photometric stream candidates cover a similar width as $N$-body models.
    We calculate it to be $w=0.80^\circ~\!{}\!{}_{-0.33^\circ}^{+0.58^\circ}$ ($1.63_{-0.68}^{+1.16}~{\rm kpc}$) from applying a mixture model accounting for the non-uniformity of our footprint to the CraII stream candidates.
    This width is consistent with both cored and cuspy models from \citet{Errani15} at its most recent pericenter.
\end{itemize}

Overall, these results strongly support evidence that CraII possesses extended tidal tails and presents new reasoning for how the width of this stream could be consistent with either a cuspy or cored dark matter halo.
Following our detection of the first photometrically confirmed RGB stars populating CraII's stellar stream, we recommend further observations to obtain radial velocities and expand the extent of the measured footprint.
Our CaHK candidates have magnitudes ($g<20.5$) that are able to be observed by current spectrographs \citep[e.g. AAT/2df/AAOmega, Magellan/IMACS,][]{Fu19, Ji2021}, making them excellent targets for spectroscopic follow-up.
Notably, we emphasize the need to cover a wider $\phi_1$ range ($|\phi_1|>8^\circ$) to better constrain our velocity dispersion and width estimates at distances from the center where disruption has settled.
We also emphasize the need for more theoretical work comparing widths and velocity dispersion in cored and cuspy dark matter models.
We still seek to explain how such a large galaxy with a low velocity dispersion formed within $\Lambda$CDM cosmology with lower mass loss than nominally needed to reproduce its properties.
Our sample of stars from the tidal tails can potentially improve $N$-body models since the one in this paper relied on properties of CraII from its center.
Understanding CraII's unusual properties is a key step to learning about diffuse galaxy formation and the challenge it may pose to dark matter models.
Now that we have found RGB stars in its tails, the CraII stream can provide important observations to advance these objectives.

\section*{Acknowledgments}

KRA acknowledges this material is based upon work supported by the National Science Foundation Graduate Research Fellowship under Grant No. 2234693. 
AC is supported by a Brinson Prize Fellowship grant through the Brinson Foundation.

WC gratefully acknowledges support from a Gruber Science Fellowship at Yale University. 
This material is based upon work supported by the National Science Foundation Graduate Research Fellowship Program under Grant No.\ DGE2139841.
The work of VMP is supported by NOIRLab, which is managed by the Association of Universities for Research in Astronomy (AURA) under a cooperative agreement with the U.S. National Science Foundation.
DJS acknowledges support from NSF grants AST-2205863 and 2508746.
BMP acknowledges support from NSF grant AST2508745.

The DELVE project is partially supported by the NASA Fermi Guest Investigator Program Cycle 9 No. 91201. 
This work is partially supported by Fermilab LDRD project L2019-011. 
This material is based upon work supported by the National Science Foundation under Grant No. AST-2108168, AST-2108169, AST-2307126, and AST-2407526.

This project used data obtained with the Dark Energy Camera (DECam), which was constructed by the Dark Energy Survey (DES) collaboration. 
Funding for the DES Projects has been provided by the US Department of Energy, the U.S. National Science Foundation, the
Ministry of Science and Education of Spain, the Science and Technology Facilities Council of the United Kingdom, the Higher Education Funding Council for England, the National Center for Supercomputing Applications at the University of Illinois at Urbana–Champaign, the Kavli Institute for Cosmological Physics at the University of Chicago, the Center for Cosmology and Astro-Particle Physics at the Ohio State University, the Mitchell Institute for Fundamental Physics and Astronomy at Texas A\&M University, Financiadora de Estudos
e Projetos, Funda\c{c}\~{a}o Carlos Chagas Filho de Amparo
\`{a} Pesquisa do Estado do Rio de Janeiro, Conselho 12 Nacional de Desenvolvimento Cient\'{ı}fico e Tecnol\'{o}gico
and the Minist\'{e}rio da Ci\^{e}ncia, Tecnologia e Inova\c{c}\~{a}o, the Deutsche Forschungsgemeinschaft and the Collaborating Institutions in the Dark Energy Survey.

The Collaborating Institutions are Argonne National Laboratory, the University of California at Santa Cruz,
the University of Cambridge, Centro de Investigaciones
En\'{e}rgeticas, Medioambientales y Tecnol\'{o}gicas–Madrid,
the University of Chicago, University College London, the DES-Brazil Consortium, the University of Edinburgh, the Eidgen{\"o}ssische Technische Hochschule (ETH) Z{\"u}rich, Fermi National Accelerator Laboratory, the University of Illinois at Urbana-Champaign, the Institut de Ci\`{e}ncies de l’Espai (IEEC/CSIC), the Institut de F\'{ı}sica d’Altes Energies, Lawrence Berkeley National Laboratory, the Ludwig-Maximilians Universit{\"a}t M{\"u}nchen and the associated Excellence Cluster Universe, the University of Michigan, NSF NOIRLab, the University of Nottingham, the Ohio State University, the OzDES Membership Consortium, the University of Pennsylvania, the University of Portsmouth, SLAC National Accelerator Laboratory, Stanford University, the University of Sussex, and Texas A\&M University.

Based on observations at NSF Cerro Tololo Inter-American Observatory, NSF NOIRLab (NOIRLab Prop. ID 2019A-0305; PI: Alex Drlica-Wagner, NOIRLab Prop. ID 2023B-646244; PI: Anirudh Chiti, NOIRLab Prop. ID 2025A-402104: PI: Andrew Pace, and NOIRLab Prop. ID 2024A-974884: PI: Andrew Pace), which is managed by the Association of Universities for Research in Astronomy (AURA) under a cooperative agreement with the U.S. National Science Foundation.

This manuscript has been authored by Fermi Research Alliance, LLC under Contract No. DE-AC02-07CH11359 with the U.S. Department of Energy, Office of Science, Office of High Energy Physics. The United States Government retains and the publisher, by accepting the article for publication, acknowledges that the United States Government retains a non-exclusive, paid-up, irrevocable, world-wide license to publish or reproduce the published form of this manuscript, or allow
others to do so, for United States Government purposes.

This work has made use of data from the European Space Agency (ESA) mission {\it Gaia} (\url{https://www.cosmos.esa.int/gaia}), processed by the {\it Gaia} Data Processing and Analysis Consortium (DPAC, \url{https://www.cosmos.esa.int/web/gaia/dpac/consortium}).
Funding for the DPAC has been provided by national institutions, in particular the institutions participating in the {\it Gaia} Multilateral Agreement.

{\it Software:} 
{\texttt{matplotlib} \citep{matplotlib},
\texttt{astropy} \citep{astropy,astropy:2018},
\texttt{numpy} \citep{numpy}, 
\texttt{scipy} \citep{scipy},  
\texttt{pandas} \citep{pandas},
\texttt{TOPCAT} \citep{topcat},
\texttt{corner.py} \citep{corner},
\texttt{galpy} \citep{galpy},
\texttt{emcee} \citep{emcee},
\texttt{scikit-learn} \citep{scikit-learn}
}

\bibliography{main}{}
\bibliographystyle{aasjournal}

\appendix

\section{Besan\c{c}on Error Fitting}\label{sec:appendix}

\begin{figure}[h!]
    \centering
    \includegraphics[width=1\linewidth]{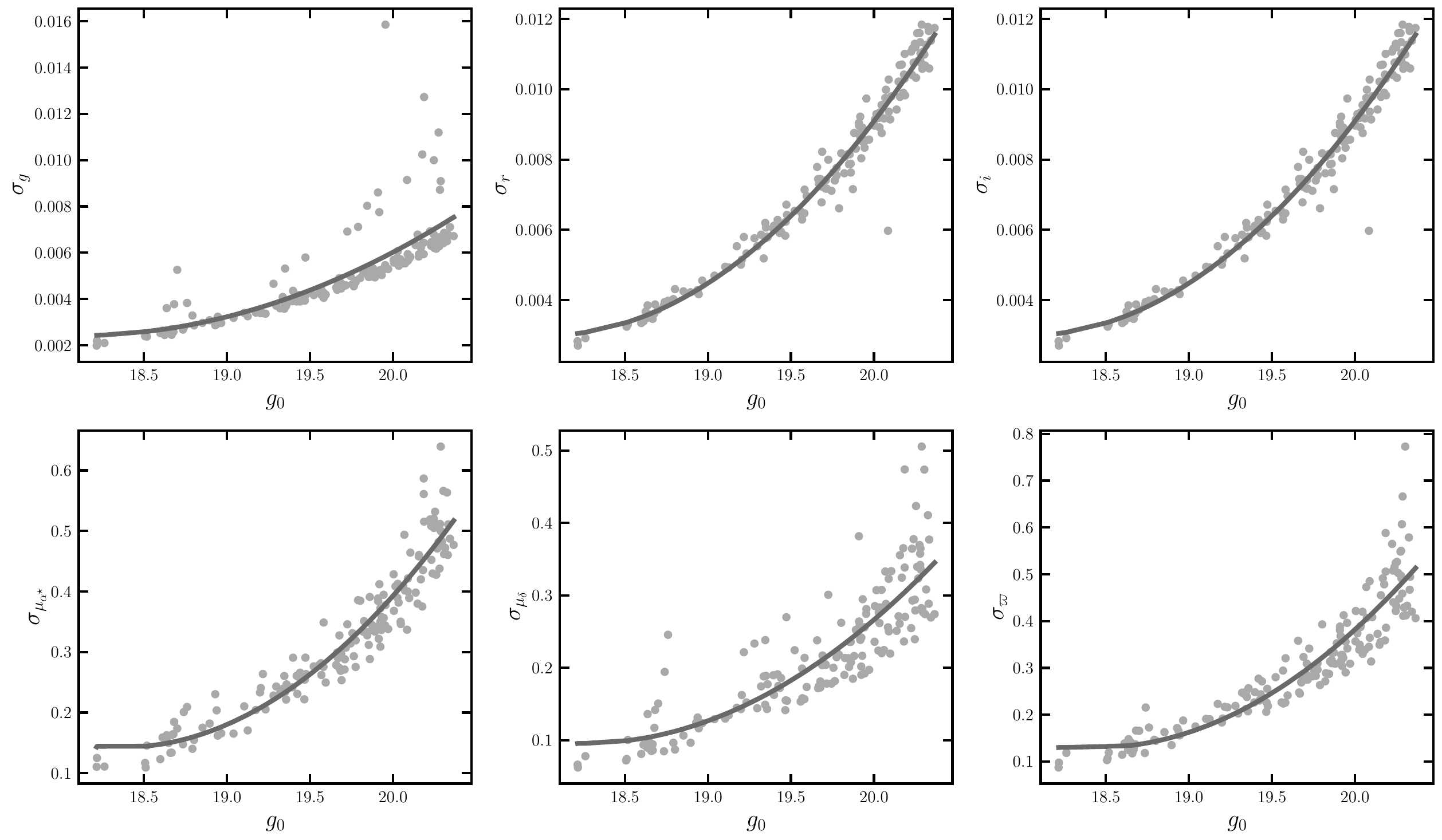}
    \caption{Error fits for magnitude, proper motion right ascension ($\mu_{\alpha^\star}$), proper motion declination ($\mu_\delta$), and parallax ($\varpi$) as a function of $g_0$ magnitude.
    These functions for the CraII MAGIC data are applied to the BMW simulations.}
    \label{fig:bmw_errs}
\end{figure}

The Besan\c{c}on catalog simulation with kinematics in the SDSS + JHK photometric system allows proper motion and parallax errors to be fixed or calculated from photometric error functions.
However, we opt to derive them from functions of CraII-specific data from the MAGIC catalog to make them more relevant for the purposes of this analysis.
A simple quadratic curve fitting function is applied to each magnitude, proper motion, and parallax error of the color-magnitude, membership score, and metallicity selected CraII stars as a function of broadband magnitudes from DELVE.
These are shown in Figure \ref{fig:bmw_errs} using the $g_0$ band magnitude.
Due to its close proximity to the Orphan-Chenab stream \citep{Belokurov07a,Casey13,Shipp18,Koposov19a,Erkal19}, some stars from pointings on the far side of the tidal tails have exceptionally low errors from deep prior surveys.
There are also a few high $g_0$ magnitudes relative to the best fit function from pointings that were not dithered.
The remaining scatter appears as expected as a function of magnitude.
Overall, this provides us with well characterized errors that can be applied to nearby Milky Way foreground.

\end{document}

%% file: authors.tex
\author[orcid=0000-0001-9649-8103]{K. R. Atzberger}
\affiliation{Department of Astronomy, University of Virginia, 530 McCormick Road, Charlottesville, VA 22904, USA}
\email[show]{katzberger@email.virginia.edu}

\author[orcid=0000-0002-6021-8760]{A. B. Pace}
\altaffiliation{Galaxy Evolution and Cosmology (GECO) Fellow}
\affiliation{Department of Astronomy, University of Virginia, 530 McCormick Road, Charlottesville, VA 22904, USA}
\email{pvpace1@gmail.com}

\author[orcid=0000-0002-3204-1742]{N. Kallivayalil}
\affiliation{Department of Astronomy, University of Virginia, 530 McCormick Road, Charlottesville, VA 22904, USA}
\email{njk3r@virginia.edu}

\author[orcid=0000-0002-7155-679X]{A. Chiti}
\affiliation{Kavli Institute for Particle Astrophysics \& Cosmology, P.O. Box 2450, Stanford University, Stanford, CA 94305, USA}
\email{achiti@stanford.edu}

\author[orcid=0000-0002-8448-5505]{D. Erkal}
\affiliation{School of Mathematics and Physics, University of Surrey, Guildford, GU2 7XH, UK}
\email{d.erkal@surrey.ac.uk}

%% added with WGR (hack sessions)

\author[orcid=0000-0003-1697-7062]{W.~Cerny}
\affiliation{Department of Astronomy, Yale University, New Haven, CT 06520, USA}
\email{william.cerny@yale.edu}

\author[orcid=0000-0002-9269-8287]{G.~Limberg}
\affiliation{Kavli Institute for Cosmological Physics, University of Chicago, Chicago, IL 60637, USA}
\email{limberg@uchicago.edu}

\author[orcid=0000-0003-4479-1265]{V.~M.~Placco}
\affiliation{NSF NOIRLab, 950 N. Cherry Ave., Tucson, AZ 85719, USA}
\email{vinicius.placco@noirlab.edu}

\author[orcid=0000-0002-8217-5626]{D.~S.~Prabhu}
\affiliation{Department of Astronomy/Steward Observatory, 933 North Cherry Avenue, Room N204, Tucson, AZ 85721-0065, USA}
\email{deepthisprabhu@gmail.com}

\author[orcid=0000-0003-1479-3059]{G.~S.~Stringfellow}
\affiliation{Center for Astrophysics and Space Astronomy, University of Colorado Boulder, Boulder, CO 80309, USA}
\email{Guy.Stringfellow@colorado.edu}

\author[orcid=0000-0003-4341-6172]{A.~K.~Vivas}
\affiliation{Cerro Tololo Inter-American Observatory/NSF NOIRLab, Casilla 603, La Serena, Chile}
\email{kathy.vivas@noirlab.edu}

%% added with WGR (all)

\author[orcid=0000-0001-5143-1255]{A.~Chaturvedi}
\affiliation{School of Mathematics and Physics, University of Surrey, Guildford, GU2 7XH, UK}
\email{aa07223@surrey.ac.uk}

\author[orcid=0000-0001-6957-1627]{P.~S.~Ferguson}
\affiliation{DiRAC Institute, Department of Astronomy, University of Washington, 3910 15th Ave NE, Seattle, WA, 98195, USA}
\email{pferguso@uw.edu}

\author[orcid=0000-0001-5805-5766]{A.~H.~Riley}
\affiliation{Lund Observatory, Division of Astrophysics, Department of Physics, Lund University, SE-221 00 Lund, Sweden}
\email{alexander.riley@fysik.lu.se}

\author[orcid=0000-0003-4102-380X]{D.~J.~Sand}
\affiliation{Department of Astronomy/Steward Observatory, 933 North Cherry Avenue, Room N204, Tucson, AZ 85721-0065, USA}
\email{dsand@arizona.edu}

%% added with CWR (comments)

\author[orcid=0000-0002-3936-9628]{J.~L.~Carlin}
\affiliation{Rubin Observatory/AURA, 950 North Cherry Avenue, Tucson, AZ, 85719, USA}
\email{jeffreylcarlin@gmail.com}

\author[orcid=0000-0003-1680-1884]{Y.~Choi}
\affiliation{NSF NOIRLab, 950 N. Cherry Ave., Tucson, AZ 85719, USA}
\email{yumi.choi@noirlab.edu}

\author[orcid=0000-0002-1763-4128]{D.~Crnojevi\'c}
\affiliation{Department of Physics \& Astronomy, University of Tampa, 401 West Kennedy Boulevard, Tampa, FL 33606, USA}
\email{dcrnojevic@ut.edu}

\author[orcid=0000-0001-8251-933X]{A.~Drlica-Wagner}
\affiliation{Fermi National Accelerator Laboratory, P.O.\ Box 500, Batavia, IL 60510, USA}
\affiliation{Kavli Institute for Cosmological Physics, University of Chicago, Chicago, IL 60637, USA}
\affiliation{Department of Astronomy and Astrophysics, University of Chicago, Chicago, IL 60637, USA}
\affiliation{NSF-Simons AI Institute for the Sky (SkAI), 172 E. Chestnut St., Chicago, IL 60611, USA}
\email{kadrlica@fnal.gov}

\author[orcid=0000-0002-4863-8842]{A.~P.~Ji}
\affiliation{Department of Astronomy \& Astrophysics, University of Chicago, 5640 S Ellis Avenue, Chicago, IL 60637, USA}
\affiliation{Kavli Institute for Cosmological Physics, University of Chicago, Chicago, IL 60637, USA}
\email{alexji@uchicago.edu}

\author[orcid=0000-0002-9110-6163]{T.~S.~Li}
\affiliation{David A. Dunlap Department of Astronomy \& Astrophysics, University of Toronto, 50 St George Street, Toronto ON M5S 3H4, Canada}
\affiliation{Department of Astronomy and Astrophysics, University of Toronto, 50 St. George Street, Toronto ON, M5S 3H4, Canada}
\email{ting.li@astro.utoronto.ca}

\author[orcid=0000-0002-9144-7726]{C.~E.~Mart\'inez-V\'azquez}
\affiliation{NSF NOIRLab, 670 N. A'ohoku Place, Hilo, Hawai'i, 96720, USA}
\email{clara.martinez@noirlab.edu}

\author[orcid=0000-0003-0105-9576]{G.~E.~Medina}
\affiliation{David A. Dunlap Department of Astronomy \& Astrophysics, University of Toronto, 50 St George Street, Toronto ON M5S 3H4, Canada}
\affiliation{Department of Astronomy and Astrophysics, University of Toronto, 50 St. George Street, Toronto ON, M5S 3H4, Canada}
\email{gustavo.medina@utoronto.ca}

\author[orcid=0000-0002-8282-469X]{N.~E.~D.~No\"el}
\affiliation{Department of Physics, University of Surrey, Guildford GU2 7XH, UK}
\email{n.noel@surrey.ac.uk}

\author[orcid=0000-0002-7123-8943]{A.~R.~Walker}
\affiliation{Cerro Tololo Inter-American Observatory/NSF NOIRLab, Casilla 603, La Serena, Chile}
\email{alistair.walker@noirlab.edu}

%% added with CWR (builders)

\author[orcid=0000-0002-3690-105X]{J.~A.~Carballo-Bello}
\affiliation{Instituto de Alta Investigaci\'on, Universidad de Tarapac\'a, Casilla 7D, Arica, Chile}
\email{jcarballo@academicos.uta.cl}

\author[orcid=0000-0001-5160-4486]{D.~J.~James}
\affiliation{ASTRAVEO LLC, PO Box 1668, Gloucester, MA 01931, USA}
\affiliation{Applied Materials Inc., 35 Dory Road, Gloucester, MA 01930, USA}
\email{djames44@gmail.com}

\author[orcid=0000-0001-9649-4815]{B.~Mutlu-Pakdil}
\affiliation{Department of Physics and Astronomy, Dartmouth College, Hanover, NH 03755, USA}
\email{Burcin.Mutlu-Pakdil@dartmouth.edu}

\author[orcid=0000-0001-9438-5228]{M.~Navabi}
\affiliation{Department of Physics, University of Surrey, Guildford GU2 7XH, UK}
\email{m.navabi@surrey.ac.uk}

\author[orcid=0000-0002-1594-1466]{J.~D.~Sakowska}
\affiliation{Department of Physics, University of Surrey, Guildford GU2 7XH, UK}
\affiliation{Instituto de Astrof\'isica de Andaluc\'ia (CSIC), Glorieta de la Astronom\'ia,  E-18080 Granada, Spain}
\email{jsakowska@iaa.es}

\collaboration{all}{MAGIC \& DELVE Collaborations}